\documentclass[useAMS,usenatbib,usegraphicx]{mn2e}

\newcommand{\Porb}{\mbox{$P_{\mathrm{orb}}$}}
\newcommand{\Teff}{\mbox{$T_{\mathrm{eff}}$}}
\newcommand{\Twd}{\mbox{$T_{\mathrm{eff,WD}}$}}
\newcommand{\Tsec}{\mbox{$T_{\mathrm{eff,sec}}$}}
\newcommand{\Msun}{\mbox{$\mathrm{M}_{\odot}$}}
\newcommand{\Rsun}{\mbox{$\mathrm{R}_{\odot}$}}
\newcommand{\Mwd}{\mbox{$M_{\mathrm{WD}}$}}
\newcommand{\Mwds}{\mbox{$M_{\mathrm{WD,spfit}}$}}
\newcommand{\Mwdl}{\mbox{$M_{\mathrm{WD,lcfit}}$}}
\newcommand{\Msec}{\mbox{$M_{\mathrm{sec}}$}}

\newcommand{\Msecl}{\mbox{$M_{\mathrm{sec,lcfit}}$}}
\newcommand{\Rwd}{\mbox{$R_{\mathrm{WD}}$}}
\newcommand{\Rsec}{\mbox{$R_{\mathrm{sec}}$}}

\newcommand{\Lines}[3]{\Ion{#1}{#2}\,$\lambda\lambda$\,#3}
\newcommand{\Ion}[2]{#1{\,\scriptsize #2}}

\newcommand{\kms}{\mbox{$\mathrm{km\,s^{-1}}$}}
\title[Four eclipsing white dwarf main sequence binaries]{Post Common
  Envelope Binaries from SDSS. V: Four eclipsing white dwarf main
  sequence binaries}

\author[S.Pyrzas et al.]{
S. Pyrzas$^{{1},{2}}$\thanks{E-mail: S.Pyrzas@warwick.ac.uk},
B. T. G\"ansicke$^{1}$, 
T. R. Marsh$^{1}$,
A. Aungwerojwit$^{{1},{3}}$,
A. Rebassa-Mansergas$^{1}$, \newauthor 
P. Rodr{\'i}guez-Gil$^{{2},{4}}$
J. Southworth$^{1}$, 
M. R. Schreiber$^{5}$, 
A. Nebot Gomez-Moran$^{6}$, and \newauthor
D.Koester$^{7}$\\ 
$^{1}$Department of Physics, University of Warwick, Coventry, CV4 7AL,
  UK\\ 
$^{2}$Isaac Newton Group of Telescopes, Apartado de correos 321, S/C de la Palma, E-38700, Canary
  Islands, Spain\\  
$^{3}$Department of Physics, Faculty of Science, Naresuan
  University, Phitsanulok 65000, Thailand\\ 
$^{4}$Instituto de Astrof{\'i}sica de Canarias, V{\'i}a L{\'a}ctea,
  s/n, La Laguna, E-38205, Tenerife,  Spain\\
$^{5}$Departamento de Fisica y Astronomia, Universidad de Valparaiso,
  Avenida Gran Bretana  1111, Valparaiso, Chile\\ 
$^{6}$Astrophysikalisches Institut Potsdam, An der Sternwarte 16, D-14482 Potsdam, Germany\\
$^{7}$Institut f\"ur Theoretische Physik und Astrophysik, University of Kiel, 24098 Kiel, Germany}
\begin{document}

\date{Accepted 2008 December 11. Received 2008 December 2; in original form 2008 November 4}

\pagerange{\pageref{firstpage}--\pageref{lastpage}} \pubyear{2002}

\maketitle

\label{firstpage}

\begin{abstract}
We identify SDSS\,011009.09+132616.1, SDSS\,030308.35+005444.1, SDSS\,143547.87+373338.5 and
SDSS\,154846.00+405728.8 as four eclipsing white dwarf plus main sequence
(WDMS) binaries from the Sloan Digital Sky Survey, and report on
follow-up observations of these systems. SDSS\,0110+1326,
SDSS\,1435+3733 and SDSS\,1548+4057 contain DA white dwarfs, while
SDSS\,0303+0054 contains a cool DC white dwarf. Orbital periods and
ephemerides have been established from multi-season
photometry. SDSS\,1435+3733, with $\Porb=3$\,h has the shortest
orbital period of all known eclipsing WDMS binaries. As for the other
systems, SDSS\,0110+1326 has $\Porb=8$\,h, SDSS\,0303+0054 has $\Porb=3.2$\,h 
and SDSS\,1548+4057 has $\Porb=4.4$\,h. Time-resolved
spectroscopic observations have been obtained and the H$\alpha$ and
\Lines{Ca}{II}{8498.02,8542.09,8662.14} triplet emission lines, as well as the 
\Lines{Na}{I}{8183.27,8194.81} absorption doublet were used to measure the 
radial velocities of the secondary
stars in all four systems. A spectral decomposition/fitting technique
was then employed to isolate the contribution of each of the
components to the total spectrum, and to determine the white dwarf
effective temperatures and surface gravities, as well as the spectral types
of the companion stars.  We used a light curve modelling code for
close binary systems to fit the eclipse profiles and the ellipsoidal
modulation/reflection effect in the light curves, to further constrain
the masses and radii of the components in all systems. All three DA
white dwarfs have masses of $\Mwd\sim0.4-0.6\,\Msun$, in line with the
expectations from close binary evolution. The DC white dwarf in
SDSS\,0303+0054 has a mass of $\Mwd\ga0.85\,\Msun$, making it
unusually massive for a post-common envelope system. The companion
stars in all four systems are M-dwarfs of spectral type M4
and later. Our new additions raise the number of known eclipsing WDMS
binaries to fourteen, and we find that the average white dwarf mass in
this sample is $<\Mwd>=0.57\pm0.16\,\Msun$, only slightly lower than the
average mass of single white dwarfs. The majority of all eclipsing
WDMS binaries contain low-mass ($<0.6\,\Msun$) secondary stars,
and will eventually provide  valuable observational input for the
calibration of the mass-radius relations of low-mass main sequence
stars and of white dwarfs.
\end{abstract}


%-------------------------------------------------------------------------------------------------------%
%-------------------------------------------------------------------------------------------------------%

\section{Introduction}
White dwarfs and low-mass stars represent the most common types of
stellar remnant and main sequence star, respectively, encountered in
our Galaxy. Yet, despite being very common, very few white dwarfs and
low mass stars have accurately determined radii and
masses. Consequently, the finite temperature mass-radius relation of white dwarfs
\citep[e.g.][]{wood95-1,paneietal00-2} remains largely untested by
observations \citep{provencaletal98-1}. In the case of low mass stars,
the empirical measurements consistently result in radii up to 15\%
larger and effective temperatures 400\,K or more below the values
predicted by theory \citep[e.g.][]{ribas06-1, lopez-morales07-1}. This
is most clearly demonstrated using low-mass eclipsing binary stars
\citep{bayless+orosz06-1}, but is also present in field stars
\citep{bergeretal06-1, moralesetal08-1} and the host stars of
transiting extra-solar planets \citep{torres07-1}.

Eclipsing binaries are the key to determine accurate stellar masses
and radii \citep[e.g.][]{andersen91-1, southworthetal07-4}. However,
because of their intrinsic faintness, very few binaries containing
white dwarfs and/or low mass stars are currently known.  Here, we
report the first results of a programme aimed at the identification of
eclipsing white dwarf plus main sequence (WDMS) binaries, which will
provide accurate empirical masses and radii for both types of
stars. Because short-period WDMS binaries underwent common envelope
evolution, they are expected to contain a wide range of white dwarf
masses, which will be important for populating the empirical white
dwarf mass-radius relation. Eclipsing WDMS binaries will also be of
key importance in filling in the mass-radius relation of low-mass
stars at masses $\la0.6\,\Msun$.

The structure of the paper is as follows. The target selection for
this programme is described in Sect.\,\ref{s-targets}. In
Sect.\,\ref{s-observations} we present our observations and data
reduction in detail. We determine the orbital periods and ephemerides
of the four eclipsing WDMS binaries in Sect.\,\ref{s-periods}, and
measure the radial velocities of the secondary stars in
Sect.\,\ref{s-k2}.  In Sect.\,\ref{s-spectralfit} we derive initial
estimates of the stellar parameters from fitting the SDSS spectroscopy
of our targets. Basic equations for the following analysis are
outlined in Sect.\,\ref{s-equations}. In Sect.\,\ref{s-lcmodels} we
describe our fits to the observed light curves, and present our
results in Sect.\,\ref{s-results}. The past and future evolution of
the four stars is explored in Sect.\,\ref{s-evolution}. Finally, we
discuss and summarise our findings, including an outlook on future
work in Sect.\,\ref{s-conclusions}.

\section{Target selection}
\label{s-targets}

We have selected eclipsing SDSS WDMS binaries based on the available
information on the radial velocities of their companion stars, and/or
evidence of a strong reflection effect. 

Initially, we used SDSS spectroscopy to measure the radial velocity of
the companion star either from the \Lines{Na}{I}{8183.27,8194.81}
absorption doublet, or from the H$\alpha$ emission line (see
\citealt{rebassa-mansergasetal07-1} for
details). SDSS\,J030308.35+005444.1 (henceforth SDSS\,0303+0054) and
SDSS\,J143547.87+373338.6 (henceforth SDSS\,1435+3733) exhibited the
largest secondary star radial velocities among $\sim1150$ WDMS
binaries which have SDSS spectra of sufficiently good quality,
287\,\kms\ and 335\,\kms, respectively.  For SDSS\,J011009.09+132616.1
(henceforth SDSS\,0110+1326) two SDSS spectra are available, which differ substantially in the
strength of the emission lines from the heated companion
star. Photometric time series (Sect.\,\ref{s-observations}) revealed
white dwarf eclipses in all three objects. SDSS\,1435+3733 has been
independently identified as an eclipsing WDMS binary by
\citet{steinfadtetal08-1}.

As the number of SDSS WDMS binaries with known orbital periods,
\Porb, and radial velocity amplitudes, $K_\mathrm{sec}$, is steadily growing
\citep{schreiberetal08-1, rebassa-mansergasetal08-1}, we are now in
the position to further pinpoint the selection of candidates for eclipses:
with \Porb\ and $K_\mathrm{sec}$ from our time-series spectroscopy, and \Mwd\ and
\Msec\ from our spectral decomposition/fitting of the SDSS spectra
\citep{rebassa-mansergasetal07-1}, we can estimate the binary
inclination from Kepler's third law. In the case of
SDSS\,J154846.00+405728.8 (henceforth SDSS\,1548+4057), the available
information suggested $i\sim85^\circ$, and time-series photometry
confirmed the high inclination through the detection of eclipses.

Full coordinates and SDSS $u, g, r, i, z$ point-spread function
magnitudes of the four systems are given in Table \ref{camg}.

\section{Observations and data reduction}
\label{s-observations}

\begin{figure}
 \includegraphics[width=84mm]{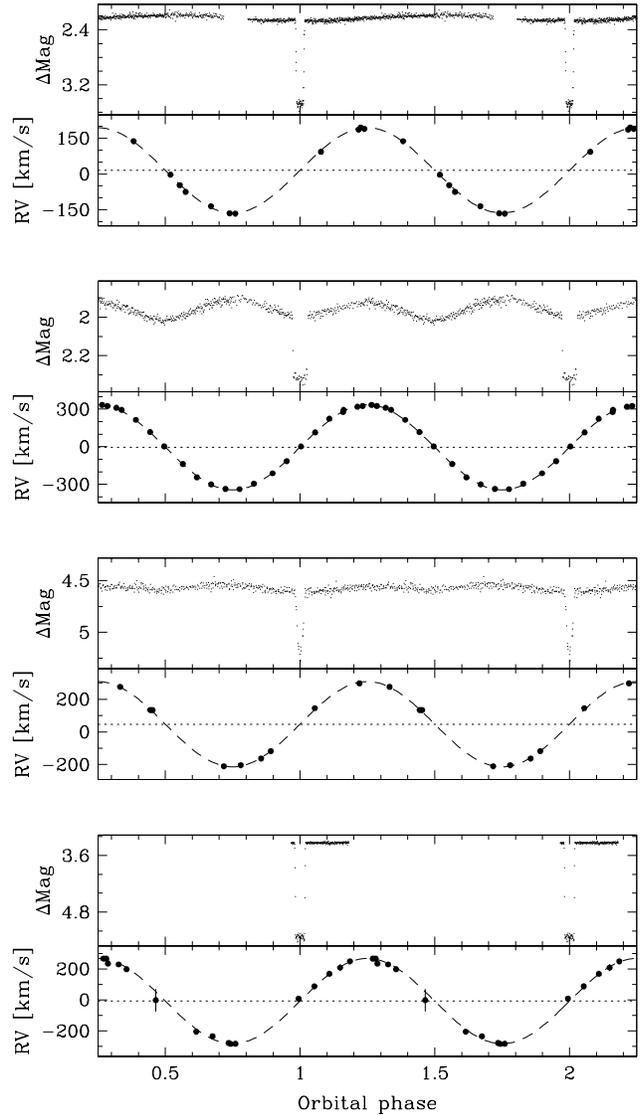}
  \caption{Phase-folded light- and radial velocity curves of the four systems. From top to bottom
    (two panels for each system): CA2.2 $I$-band light curve and
    \Ion{Ca}{II} radial velocity curve of SDSS\,0110+1326, CA2.2 fitlerless
    light curve and \Ion{Na}{I} radial velocity curve of
    SDSS\,0303+0054, IAC80 $I$-band light curve and \Ion{Na}{I} radial
    velocity curve of SDSS\,1435+3733 and WHT $R$-band light curve and
    \Ion{Na}{I} radial velocity curve of SDSS\,1548+4057.}
  \label{all}
\end{figure}

Follow-up observations of the four systems - time series CCD
photometry and time-resolved spectroscopy - were obtained at five
different telescopes, namely the 4.2m William Herschel Telescope
(WHT), the 2.5m Nordic Optical Telescope (NOT), the 2.2m telescope in
Calar Alto (CA2.2), the 1.2m Mercator telescope (MER) and the IAC 0.8m
(IAC80). A log of the observations is given in Table\,\ref{log}, while
Fig.\,\ref{all} shows phase-folded light curves and radial velocity
curves of all the systems. A brief account of the used instrumentation
and the data reduction procedures is given below.

\begin{table}
\centering
\caption{Full SDSS names and $u,g,r,i,z$ magnitudes of
   SDSS\,0110+1326, SDSS\,0303+0054, SDSS\,1435+3733 and SDSS\,1548+4057.}
\label{camg}
\begin{tabular}{lccccc}
\hline SDSS\,J & $u$ & $g$ & $r$ & $i$ & $z$ \\ 
\hline 
011009.09+132616.1 & 16.51 & 16.53 & 16.86 & 17.02 & 16.94 \\ 
030308.35+005444.1 & 19.14 & 18.60 & 18.06 & 16.89 & 16.04 \\ 
143547.87+373338.5 & 17.65 & 17.14 & 17.25 & 16.98 & 16.66 \\ 
154846.00+405728.8 & 18.79 & 18.32 & 18.41 & 18.17 & 17.68 \\
\hline
\end{tabular}
\end{table}

\begin{table}
\centering
\setlength{\tabcolsep}{0.9ex}
\caption{Log of the observations.}
\label{log}
\begin{tabular}{@{}lccccc@{}}
\hline 
Date & Obs. & Filter/Grating & Exp.\,[s] & Frames & Eclipses \\ 
\hline 
\multicolumn{6}{l}{\textbf{SDSS\,0110+1326}} \\ 
2006 Aug 04 & IAC80 & $I$ & 420    & 25 & 0 \\ 
2006 Aug 05 & IAC80 & $I$ & 60     & 170 & 0 \\ 
2006 Aug 10 & IAC80 & $I$ & 420    & 13 & 0 \\
2006 Aug 15 & IAC80 & $I$ & 420    & 15 & 0 \\
2006 Aug 16 & IAC80 & $I$ & 420    & 12 & 0 \\
2006 Sep 15 & CA2.2 & $I$ & 60     & 130 & 1 \\
2006 Sep 16 & CA2.2 & $I$ & 45--55 & 312 & 1 \\
2006 Sep 17 & CA2.2 & $I$ & 25--60 & 582 & 1 \\ 
2006 Sep 26 & WHT & R600B/R316R    & 1200 & 2 \\
2006 Sep 27 & WHT & R600B/R316R    & 600 & 4 \\
2006 Sep 29 & WHT & R600B/R316R    & 600 & 2 \\
2007 Aug 20 & CA2.2 & $BV$ & 25    & 159 & 0 \\
2007 Aug 21 & CA2.2 & $BV$ & 25    & 101 & 0 \\
2007 Sep 03 & WHT & R1200B/R600R   & 1000 & 2 \\
2007 Sep 04 & WHT & R1200B/R600R   & 1000 & 4 \\
2007 Oct 09 & MER & clear & 40     & 150 & 1 \\
\multicolumn{6}{l}{\textbf{SDSS\,0303+0054}} \\
2006 Sep 12 & CA2.2 & clear & 15--35 & 443 & 1 \\
2006 Sep 14 & CA2.2 & clear & 45--60 & 63 & 1 \\
2006 Sep 15 & CA2.2 & clear & 60     & 44 & 1 \\
2006 Sep 18 & CA2.2 & $R$   & 50--60 & 165 & 1 \\
2006 Sep 26 & WHT & R600B/R316R & 600 & 4 \\
2006 Sep 27 & WHT & R600B/R316R & 600 & 18 \\
2007 Aug 22 & CA2.2 & $BV$  & 30     & 63 & 1 \\
2007 Aug 26 & CA2.2 & $BV$  & 35     & 84 & 1 \\
2007 Oct 15 & MER & clear & 55 & 60 & 1 \\
\multicolumn{6}{l}{\textbf{SDSS\,1435+3733}} \\
2006 Jul 04 & WHT & R1200B/R600R & 720 & 1 \\
2006 Jul 05 & WHT & R1200B/R600R & 900 & 1 \\
2007 Feb 16 & IAC80 & $I$ & 90 & 45 & 1 \\
2007 Feb 17 & IAC80 & $I$ & 90 & 162 & 1 \\
2007 Feb 18 & IAC80 & $I$ & 70 & 226 & 2 \\
2007 May 18 & CA2.2 & $V$ & 15 &   9 & 1 \\
2007 May 19 & CA2.2 & $BV$ & 12 & 48 & 1 \\
2007 May 19 & CA2.2 & clear & 12-15 & 27 & 1 \\ 
2007 Jun 23 & WHT & R1200B/R600R & 1200 & 5 \\
2007 Jun 24 & WHT & R1200B/R600R & 1200 & 2 \\
\multicolumn{6}{l}{\textbf{SDSS\,1548+4057}} \\
2006 Jul 02 & WHT & R1200B/R600R & 1200 & 1 &  \\	
2006 Jul 03 & WHT & R1200B/R600R & 1500 & 1 &  \\	
2007 Jun 19 & WHT & R1200B/R600R & 1200 & 3 & \\
2007 Jun 20 & WHT & R1200B/R600R & 1200 & 2 & \\
2007 Jun 21 & WHT & R1200B/R600R & 1200 & 4 & \\
2007 Jun 22 & WHT & R1200B/R600R & 1200 & 4 & \\
2007 Jun 23 & WHT & R1200B/R600R & 1200 & 1 & \\
2007 Jun 24 & WHT & R1200B/R600R & 1200 & 2 & \\
2008 May 08 & IAC80 & $V$ & 300 & 71 & 1 \\
2008 May 10 & IAC80 & $R$ & 300 & 64 & 2 \\
2008 May 12 & IAC80 & $R$ & 300 & 5 & 1 \\
2008 Jun 26 & NOT & clear & 140 & 60  & 1 \\
2008 Jun 29 & NOT & clear & 30 & 90  & 1 \\
2008 Jul 05 & WHT & $R$ & 5 & 247 & 1 \\
\hline
\end{tabular}
\end{table}

\subsection{Photometry}

Photometric observations of the four targets were obtained at all five telescopes. In every telescope set-up, care has been taken to ensure that at least 3 good comparison stars were available in the science images, especially in the cases when the CCDs were windowed.

\subsubsection{WHT 4.2m}

The observations were carried out using the AUX-port imager, equipped
with the default 2148~x~4200 pixel E2V CCD44-82 detector, with an unvignetted, circular field-of-view (FOV) of $2.2'$ in diameter. For the observations, the CCD was binned (4~x~4) to reduce readout time to
4\,s. SDSS\,1548+4057 was observed with a Johnson $R$ filter. The
images were de-biased and flat-fielded within MIDAS and aperture
photometry was carried out using SEXTRACTOR
\citep{bertin+arnouts96-1}. A full account of the employed reduction
pipeline is given by \citet{gaensickeetal04-1}.

\subsubsection{NOT 2.5m}

The observations were carried out using the Andalucia Faint Object
Spectrograph and Camera (ALFOSC), equipped with a 2k~x~2k pixel E2V
CCD42-40 chip, with a FOV of $6.5'$~x~$6.5'$. The CCD was binned (2~x~2) to reduce readout
time. Filterless observations were obtained for SDSS\,1548+4057. The
reduction procedure for the NOT data was the same as the one used for
the WHT data.

\subsubsection{Calar Alto 2.2m}

The observations were carried out using the Calar Alto Faint Object
Spectrograph (CAFOS) equipped with the standard 2k~x~2k pixel SITe
CCD (FOV $16'$~x~$16'$). For the observations, the CCD was windowed to reduce readout
time to 10\,s. Observations were carried out using Johnson $V, R, I$, 
and ``R\"oser'' $BV$\footnote{A BG39/3 filter centred on
  $4977$\,\AA\ with a full width at half maximum (FWHM) of $1559$\,\AA} filters, as well as
without any filter. In detail, we used $I$ and $BV$ for SDSS\,0110+1326, $R$
and $BV$ for SDSS\,0303+0054, and $V$, $BV$, and no filter for
SDSS\,1435+3733. The reduction procedure was the same as described
above.

\subsubsection{Mercator 1.2m}

Filterless photometric observations have been obtained for
SDSS\,0110+1326 and SDSS\,0303+0054 using the MERcator Optical
Photometric imagEr (MEROPE) equipped with a 2k~x~2k EEV CCD
chip (FOV $6.5'$~x~$6.5'$). For the observations, the CCD was binned (3~x~3) to reduce readout time to about 8\,s. The reduction procedure for the Mercator
data was the same as above.

\subsubsection{IAC\,0.8m}

Johnson $I$-band photometry has been obtained at the IAC\,0.8m
telescope for SDSS\,0110+1326 and SDSS\,1435+3733, while $V$- and
$R$-band photometry has been obtained for SDSS\,1548+4057. The
telescope was equipped with the SI 2k~x~2k E2V CCD (FOV $10.25'$~x~$10.25'$), 
which was binned (2~x~2)
and windowed to reduce readout time to 11\,s. Data
reduction was performed within \textsf{IRAF}\footnote{IRAF is
  distributed by the National Optical Astronomy Observatory, which is
  operated by the Association of Universities for Research in
  Astronomy, Inc., under contract with the National Science
  Foundation, http://iraf.noao.edu}. After bias and flat-field
corrections the images were aligned and instrumental magnitudes were
measured by means of aperture photometry.

\subsection{Spectroscopy}

All spectroscopy presented in this work was acquired using the William
Herschel Telescope (WHT) and the Intermediate dispersion Spectrograph and
Imaging System (ISIS). For the red arm the grating used was either
R316R or R600R, and for the blue arm the grating was either R600B or
R1200B. For all observations the blue-arm detector was an EEV
2k$\times$4k CCD. In 2006 July and September the red-arm detector was
a Marconi 2k$\times$4k CCD, and in later observing runs a
high-efficiency RED+ 2k$\times$4k CCD was used. In all cases the CCDs
were binned spectrally by a factor of 2 and spatially by factors of
2--4, to reduce readout noise, and windowed in the spatial direction
to decrease the readout time.

\begin{table*}
\begin{minipage}{165mm}
\centering
\caption{Mid-eclipse timings, cycle number and the difference between
  observed and computed eclipse times using the ephemerides provided
  in equations \ref{eph1}-\ref{eph4}. The large $O-C$ values for the
  first five eclipses of SDSS\,1548+4057 are due to the poor time
  resolution of the IAC\,80 light curves (Table\,\ref{log}). Mid-eclipse
  times for SDSS\,1435+3733 from \citet{steinfadtetal08-1} are also included.}
\label{midtimes}
\begin{tabular}{@{}llrrllrr@{}}
\hline 
System & $T_{0}$ [HJD] & Cycle & $O-C$ [s] & System & $T_{0}$ [HJD] & Cycle & $O-C$ [s] \\ 
\hline
\textbf{SDSS\,0110+1326} & 2453994.447919 & 0 & +4 & & 2454150.71397 & 16 & +22 \\
                         & 2453995.445757 & 3 & -16 & & 2454239.40916 & 722 & -6 \\
                         & 2453996.444139 & 6 & +12 & & 2454240.41415 & 730 & -10 \\
                         & 2454383.692048 & 1170 & 0 & & 2454240.66553 & 732 & -1 \\
\textbf{SDSS\,0303+0054} & 2453991.616630 & 0 & +20 & & 2454249.71103 & 804 & +5 \\
                         & 2453993.498445 & 14 & -1 & & 2454251.72113 & 820 & +6 \\
                         & 2453994.708508 & 23 & -2 & & 2454252.85179 & 829 & +4 \\ 
                         & 2453997.531505 & 44 & -12 & \textbf{SDSS\,1548+4057} & 2454592.57222 & 0 & 71 \\
                         & 2454339.675049 & 2559 & -12 & & 2454597.39591 & 26 & 92 \\
                         & 2454335.642243 & 2589 & -26 & & 2454597.57771 & 27 & -230 \\
                         & 2454389.55246  & 2960 & +33 & & 2454599.62180 & 38 & 185 \\
\textbf{SDSS\,1435+3733} & 2454148.70346 & 0 & -13 & & 2454644.51613 & 280 & -11 \\
                         & 2454149.70865 & 8 & -1 & & 2454647.48471 & 296 & 15 \\
                         & 2454150.58802 & 15 & -5 & & 2454653.42122 & 328 & 11 \\
\hline
\end{tabular}
\end{minipage}
\end{table*}

Data reduction was undertaken using optimal extraction
\citep{horne86-1} as implemented in the {\sc pamela}\footnote{{\sc
    pamela} and {\sc molly} were written by TRM and can be found at
  {\tt http://www.warwick.ac.uk/go/trmarsh}} code \citep{marsh89-1},
which also makes use of the {\sc starlink}\footnote{The Starlink
  Software can be found at {\tt http://starlink.jach.hawaii.edu/}} 
packages {\sc figaro} and {\sc
  kappa}. Telluric lines removal and flux-calibration was performed
separately for each night, using observations of flux standard stars.

Wavelength calibrations were obtained in a standard fashion using
spectra of copper-argon and copper-neon arc lamps. For SDSS\,0303, arc lamp exposures
were obtained during the spectroscopic observations and the wavelength
solutions were interpolated from the two arc spectra bracketing each
spectrum. For the other objects we did not obtain dedicated arc
spectra, to increase the time efficiency of the observations. The
spectra were wavelength-calibrated using arc exposures taken at the
beginning of each night, and drift in the wavelength solution was
removed using measurements of the positions of the $\lambda$7913 and
$\lambda$6300 night-sky emission lines. This procedure has been found
to work well for spectra at red wavelengths (see \citealt{southworthetal06-1,southworthetal08-2}). 
In the blue arm, a reliable correction of the wavelength zero-point was not possible as
only one (moreover weak) sky-line (Hg\,I $\lambda$4358) is available,
and consequently, these spectra are not suitable for
velocity measurements. The reciprocal dispersion and resolution for the R316R
grating are approximately 1.7\,\AA\,px$^{-1}$ and 3.3\,\AA, and for
the R600R grating are 0.89\,\AA\,px$^{-1}$ and 1.5\,\AA, respectively.

%-------------------------------------------------------------------------------------------------------%
%-------------------------------------------------------------------------------------------------------%

\section{Orbital periods and ephemerides}
\label{s-periods}

Mid-eclipse times for all four systems were measured from their light
curves as follows. The observed eclipse profile was mirrored in time
around an estimate of the eclipse centre. The mirrored profile was
then overplotted on the original eclipse profile and shifted against
it, until the best overlap between the points during ingress and
egress was found. Given the sharpness of eclipse in/egress features,
we (conservatively) estimate the error in the mid-eclipse times to be comparable to the
duty cycle (exposure plus readout time) of the observations. Table
\ref{midtimes} lists the mid-eclipse times. An initial estimate of the
cycle count was then obtained by fitting eclipse phases
$(\phi^{\rm{observed}}_{\rm{0}} - \phi^{\rm{fit}}_{\rm{0}})^{-2}$ over
a wide range of trial periods. Once an unambiguous cycle count was
established, a linear eclipse ephemeris was fitted to the times of
mid-eclipse. For SDSS\,1435+3733 we have also used the mid-eclipse
times provided by \citet{steinfadtetal08-1}.  The resulting
ephemerides, with numbers in parenthesis indicating the error on the last digit, are

\begin{equation}
\label{eph1}
T_{\rm{0}}(\rm{HJD}) = 2453994.44787(9) + 0.3326873(1) E
\end{equation} 
\noindent
for SDSS\,0110+1326, that is, $P_{\rm{orb}} = 7.984495(3)$h

\begin{equation}
\label{eph2}
T_{\rm{0}}(\rm{HJD}) = 2453991.6164(1) + 0.13443772(7) E
\end{equation} 
\noindent
for SDSS\,0303+0054, that is, $P_{\rm{orb}} = 3.226505(1)$h 

\begin{equation}
\label{eph3}
T_{\rm{0}}(\rm{HJD}) = 2454148.70361(6) + 0.1256311(1) E
\end{equation}
\noindent
for SDSS\,1435+3733, that is, $P_{\rm{orb}} = 3.015144(2)$h and

\begin{equation}
\label{eph4}
T_{\rm{0}}(\rm{HJD}) = 2454592.57135(6) + 0.18551774(4) E
\end{equation}
\noindent
for SDSS\,1548+4057, that is, $P_{\rm{orb}} = 4.45242576(4)$h.

These ephemerides were then used to fold both the photometric and the
spectroscopic data over phase (Fig.\,\ref{all}).
%-------------------------------------------------------------------------------------------------------%
%-------------------------------------------------------------------------------------------------------%

\section{Radial velocities of the secondary stars}
\label{s-k2}

\begin{figure*}
 \includegraphics[angle=-90,width=140mm]{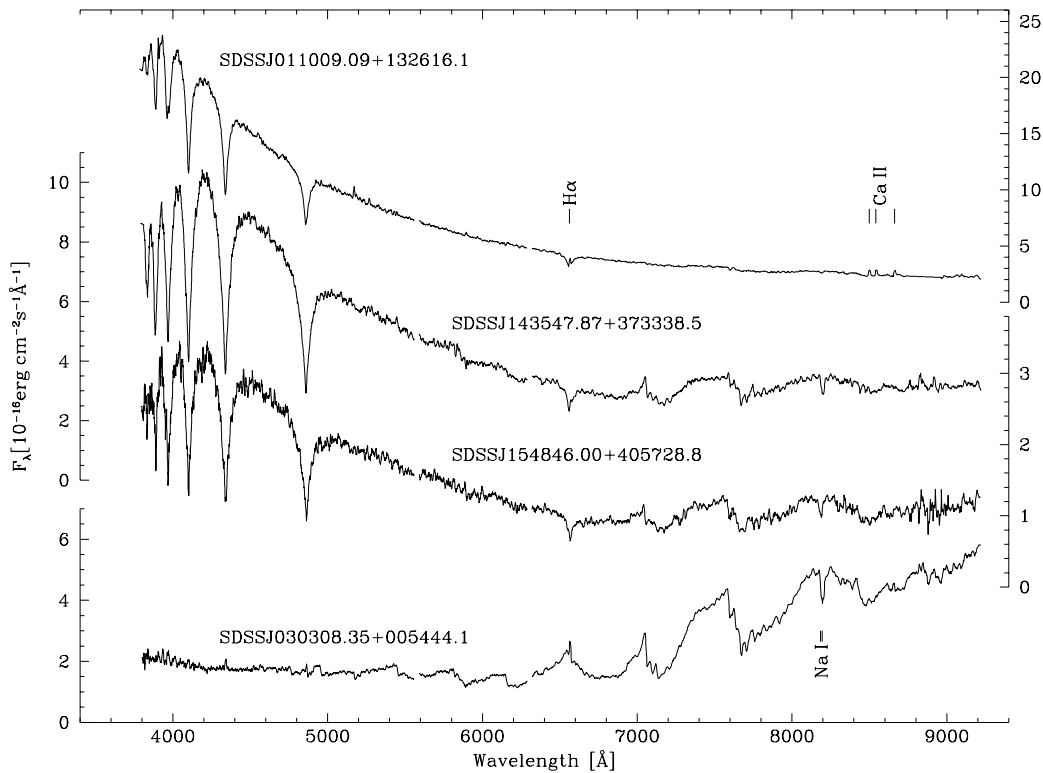}
  \caption{SDSS spectra of the four systems. Top to bottom:
    SDSS\,0110+1326, SDSS\,1435+3733 SDSS\,1548+4057 and
    SDSS\,0303+0054. The radial velocity of the companion star in SDSS0110+1326 was
measured from the $\mathrm{H}\alpha$ and \Ion{Ca}{II} emission lines, whereas for the 
other three systems, we used the \Ion{Na}{I} absorption doublet}
  \label{spec} 
\end{figure*}

\begin{figure}
 \includegraphics[width=84mm]{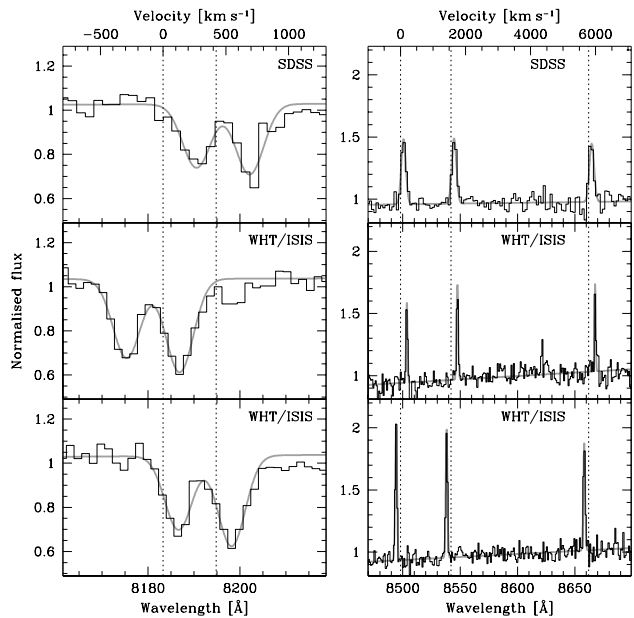}
  \caption{Sample line profile fits (gray) to the observed spectra
    (black) of SDSS\,0303+0054 (left panel,
    \Lines{Na}{I}{8183.27,8194.81} absorption doublet) and of
    SDSS\,0110+1326 (right panel, \Ion{Ca}{II} emission triplet). The
    rest wavelengths are indicated by the dotted vertical lines. The velocity
    scale at the top is relative to the blue-most component of each multiplet.
    The top spectra are from SDSS, the middle and bottom spectra were
    obtained at the WHT. Radial velocity variations in these systems
    are obvious to the eye.}
  \label{rvfit} 
\end{figure}

Figure \ref{spec} shows the SDSS spectra of our four
targets. Measuring radial velocity variations of the stellar
components needs sharp spectral features. Determining the radial
velocity amplitude of the white dwarf, $K_\mathrm{WD}$, is notoriously
difficult because of the width of the Balmer lines, and our
spectroscopic data are insufficient in quantity and signal-to-noise ratio
for the approach outlined e.g. by
\citet{maxtedetal04-1}\footnote{Ultraviolet spectroscopy allows
  accurate measurements of $K_\mathrm{WD}$ using narrow metal lines
  originating in the white dwarf photosphere, \citep{obrienetal01-1,
    odonoghueetal03-1, kawkaetal07-1}, but at the cost of space-based
  observations.}. We hence restrict our analysis to the measurement of
the radial velocity amplitude of the secondary star, $K_\mathrm{sec}$.

\subsection{SDSS\,0110+1326}

The spectrum of SDSS\,0110+1326 displays the
\Lines{Ca}{II}{8498.02, 8542.09,8662.14} emission triplet as well as an
$\mathrm{H}\alpha$ emission line. We determined the velocities of the
H$\alpha$ emission line by fitting a second-order polynomial plus a
Gaussian emission line to the spectra. For the \Ion{Ca}{II} triplet,
we fitted a second-order polynomial plus three Gaussian emission lines
with identical width and whose separations were fixed to the
corresponding laboratory values (see Fig.\,\ref{rvfit}, right panel).
The H$\alpha$ and \Ion{Ca}{II} radial velocities were then separately
phase-folded using the ephemeris Eq.\,(\ref{eph1}), and fitted with a
sine wave. The phasing of the radial velocity curves agreed with that
expected from an eclipsing binary (i.e. red-to-blue crossing at
orbital phase zero) within the errors. The resulting radial velocity amplitudes are
$K_{\mathrm{sec\,,H}\alpha}=200.1\pm4.8\,\kms$ with a systemic
velocity of $\gamma_{\mathrm{H}\alpha}=19.4\pm4.1\,\kms$ for the
$\mathrm{H}\alpha$ line; and
$K_{\mathrm{sec,\Ion{Ca}{II}}}=178.8\pm2.4\,\kms$ and
$\gamma_{\mathrm{\Ion{Ca}{II}}}=15.2\pm2.4\,\kms$ for the \Ion{Ca}{II}
line (see also Fig.\,\ref{all}). 

\begin{figure}
 \includegraphics[width=84mm]{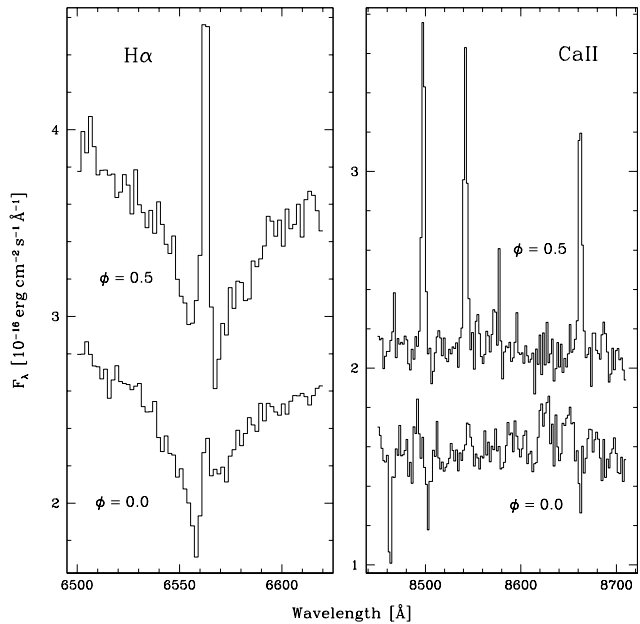}
  \caption{Average spectra of SDSS\,0110+1326. Left panel: H$\alpha$ line, right panel: \Ion{Ca}{II} triplet. The binary phase of each spectrum is clearly marked. Prior to averaging, each spectrum was shifted to the restframe of the secondary, using the measured $K_\mathrm{sec}$ and $\gamma$ values. For each line and binary phase, three spectra were averaged.}
  \label{strg} 
\end{figure}

We decided, following a suggestion by the referee,to investigate whether the emission lines originate predominantly on the illuminated hemisphere of the secondary star. If that is the case -- keeping in mind that we see the system almost edge-on -- we expect a significant variation of the line strength with orbital phase, reaching a maximum around phase $\phi=0.5$ (superior conjunction of the secondary) and almost disappearing around $\phi=0.0$. Figure\,\ref{strg} shows average spectra of SDSS\,0110+1326, focused on the H$\alpha$ line and the \Ion{Ca}{II} triplet for $\phi=0.0$ and $\phi=0.5$. It is apparent that the H$\alpha$ and \Ion{Ca}{II} emission lines are very strong near $\phi=0.5$. The H$\alpha$ emission line is very weak near $\phi=0.0$, and \Ion{Ca}{II} is seen in absorption. The equivalent width (EW) of the blue-most component of the \Ion{Ca}{II} triplet is shown in Fig.\,\ref{ew4} (top panel) as a function of orbital phase. Given the spectral resolution and quality of our data, measuring the EW of the H$\alpha$ emission line is prone to substantial uncertainties, as it is embedded in the broad H$\alpha$ absorption of the white dwarf photosphere (see Fig.\,\ref{spec} again), but it generally follows a simlilar pattern as the one seen in the \Ion{Ca}{II} triplet. This analysis supports our assumption that the emission lines originate on the irradiated, inner hemisphere of the secondary star. Hence, the centre of light of the secondary star is displaced towards the Lagrangian point $\mathrm{L}_{1}$, with respect to the centre of mass. The emission lines trace, therefore, the movement of the centre of light and not the centre of mass and as a result, the $K_\mathrm{sec}$ values measured from either the H$\alpha$ or the \Ion{Ca}{II} emission lines are very unlikely to represent the true radial velocity amplitude of the secondary star,
but instead give a lower limit to it.

The fact that $K_\mathrm{sec}$ measured from the $\mathrm{H}\alpha$ line is larger than that from the \Ion{Ca}{II} line suggests that the $\mathrm{H}\alpha$ emission is distributed in a slightly more homogeneous fashion on the secondary, and illustrates that correcting the observed $K_\mathrm{sec}$ for the effect of irradiation is not a trivial matter. We describe our approach to this issue in
Sect.\,\ref{s-lcmodels0110}. 

\subsection{SDSS\,0303+0054, SDSS\,1435+3733, and SDSS\,1548+4057}

The \Lines{Na}{I}{8183.27,8194.81} absorption doublet is a strong
feature in the spectra of SDSS\,0303+0054, SDSS\,1435+3733, and
SDSS\,1548+4057. We measured the radial velocity variation of the
secondary star in SDSS\,0303+0054 by fitting this doublet with a second-order polynomial
plus two Gaussian emission lines of common width and a separation
fixed to the corresponding laboratory value (see Fig.\,\ref{rvfit},
left panel). The same method was applied to the spectra of SDSS\,1435+3733 
and SDSS\,1548+4057. A sine-fit to each of the radial velocities data sets, 
phase-folded using the ephemerides Eq.\,(\ref{eph2})-(\ref{eph4}), gives
the radial velocity of the secondary star, $K_\mathrm{sec}$, and the systemic
velocity $\gamma$ for each system. The results of the radial velocity 
measurements are summarised in Table \ref{k2res}.

\begin{figure}
 \includegraphics[width=84mm]{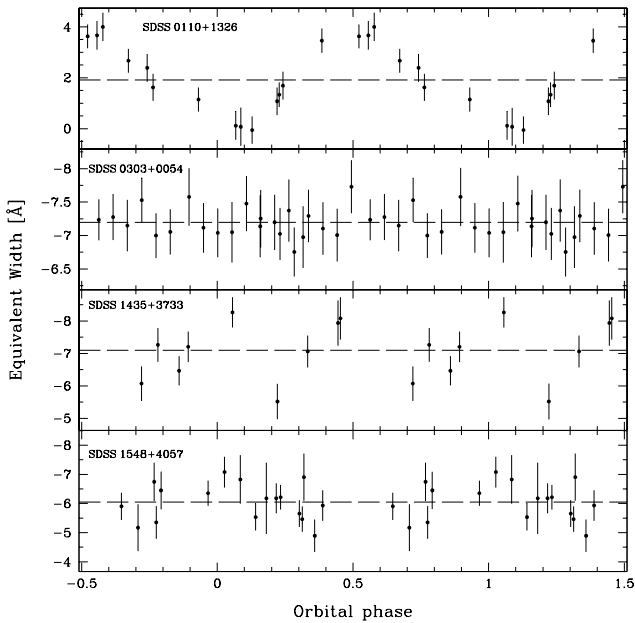}
  \caption{Equivalent widths of the lines used to measure radial velocities in all four systems with respect to the orbital phase. Top panel: blue-most component of the \Ion{Ca}{II} emission triplet in SDSS\,0110+1326. Lower three panels: blue-most component of the \Ion{Na}{I} absorption doublet in SDSS\,0303+0054, SDSS\,1435+3733 and SDSS\,1548+4057 respectively. The dashed lines indicate the
  mean EW. The strong variation of the EW of the \Ion{Ca}{II} emission line is obvious. No significant variation as a function of phase is observed for the other three systems. A full orbital cycle has been duplicated for clarity.}
  \label{ew4} 
\end{figure}

No significant variation in the strength of the \Ion{Na}{I} doublet
was observed as a function of orbital phase, and we hence assume that
$K_\mathrm{sec}$ measured from this doublet reflects the true radial
velocity amplitudes of the secondary stars in SDSS\,0303+0054,
SDSS\,1435+3733, and SDSS\,1548+4057. This is illustrated in Fig.\,\ref{ew4}, where
we plot the equivalent widths against orbital phase for these three systems.

\begin{table}
\centering 
\caption{Summary of the radial velocity measurements for all four systems.}
\label{k2res}
\begin{tabular}{@{}cccc@{}}
\hline
 System & Line & $K_{\mathrm{sec}}$ [\kms] & $\gamma$ [\kms] \\
\hline
SDSS\,0110+1326 & $\mathrm{H}\alpha$ & $200.1\pm4.8$ & $19.4\pm4.1$ \\
                & \Ion{Ca}{II} & $178.8\pm2.4$ & $15.2\pm2.4$ \\
SDSS\,0303+0054& \Ion{Na}{I} & $339.7\pm1.9$ & $-4.0\pm1.4$ \\
SDSS\,1435+3733& \Ion{Na}{I} & $260.9\pm2.9$ & $47.4\pm2.2$ \\
SDSS\,1548+4057 & \Ion{Na}{I} & $274.7\pm2.6$ & $-7.4\pm2.2$ \\
\hline
\end{tabular}
\end{table}

%-------------------------------------------------------------------------------------------------------%
%-------------------------------------------------------------------------------------------------------%

\section{Spectroscopic stellar parameters}
\label{s-spectralfit}

\begin{figure}
 \includegraphics[angle=-90,width=84mm]{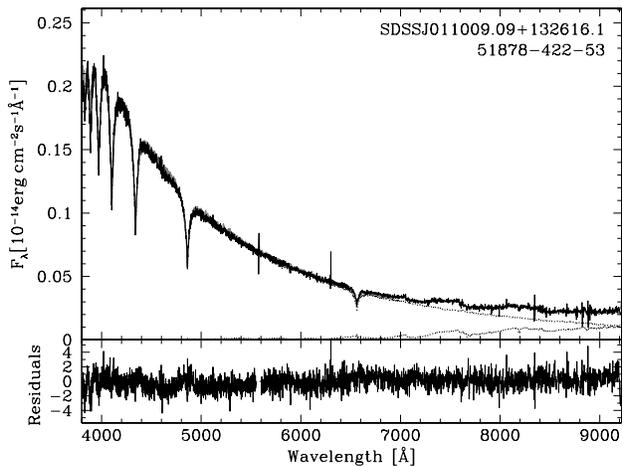}
  \caption{Two-component fit to the spectrum of SDSS\,0110+1326. The
    top panel shows the spectrum of the object as a solid black line
    and the two templates, white dwarf and M-dwarf, as dotted
    lines. The bottom panel shows the residuals from the fit.}
  \label{rdfits} 
\end{figure}

\begin{figure}
 \includegraphics[width=84mm]{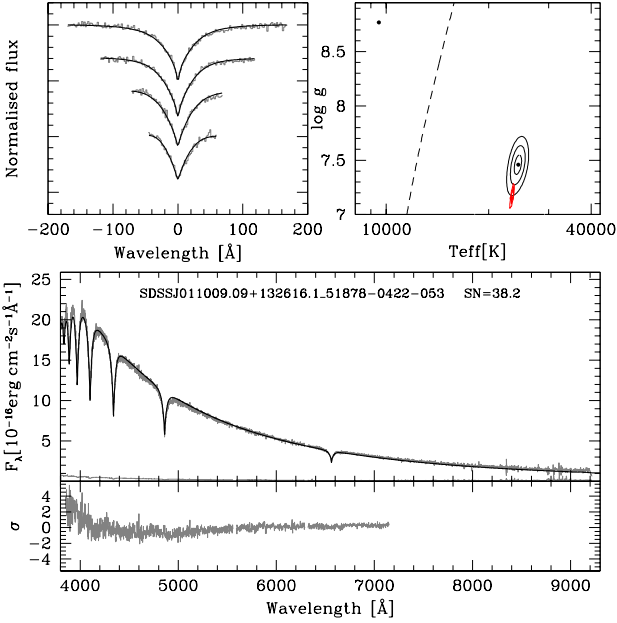}
  \caption{Spectral model fit to the white dwarf in SDSS\,0110+1326,
    obtained after subtracting the best-fit M-dwarf template. Top left
    panel: best-fit (black lines) to the observed $\mathrm{H}\beta$ to
    $\mathrm{H}\epsilon$ (gray lines, top to bottom) line
    profiles. The model spectra and observations have been normalised
    in the same way. Top right panel: 1, 2 and 3\,$\sigma$ contour
    plots in the \Teff-$\log g$ plane. The black contours refer to the
    best line profile fit, the red ones (which collapse into a dot on
    the scale of the plot) to the fit of the spectral range 3850--7150\,\AA. The
    dashed line indicates the occurrence of maximum $\mathrm{H}\beta$
    equivalent width. The best ``hot'' and ``cold'' line profile
    solutions are indicated by black dots, while the best fit to the
    whole spectrum by a red one. Bottom panel: the residual white
    dwarf spectrum resulting from the spectral decomposition and their
    flux errors (gray lines) along with the best-fit white dwarf model
    (black line) in the 3850--7150\,\AA\ wavelength range (top) and the
    residuals of the fit (gray line, bottom).}
  \label{wdfits} 
\end{figure}

We used the spectral decomposition/fitting method described in detail
in \citet{rebassa-mansergasetal07-1} to estimate the white dwarf
effective temperatures (\Teff) and surface gravities ($\log g$), as
well as the spectral types of the companion stars, for our four
targets from their SDSS spectroscopy. 

Briefly, the method employed is the following: as a first step a
two-component model is fitted to the WDMS binary spectrum using a grid
of observed M-dwarf and white dwarf templates. This step determines
the spectral type and flux contribution of the M-dwarf component as
shown in Fig.\,\ref{rdfits}.  After subtracting the best-fit M-dwarf,
the residual white dwarf spectrum is fitted with a grid of white dwarf
model spectra from \citet{koesteretal05-1}. We fit the normalised
H$\beta$ to H$\epsilon$ line profiles, omitting $\mathrm{H}\alpha$
which is most severely contaminated by the continuum and/or H$\alpha$
emission from the companion star. Balmer line profile fits can lead to
degeneracy in the determination of \Teff\, and $\log g$, as their
equivalent widths (EWs) go through a maximum at
$\Teff\simeq13000\,\mathrm{K}$, which means that fits of similar
quality can be achieved for a ``hot'' and ``cold'' solution. In order
to select the physically correct solution, we also fit the continuum
plus Balmer lines over the range 3850--7150\,\AA. The resulting
\Teff\ and $\mathrm{log}g$ are less accurate than those from line
profile fits, but  sensitive to the slope of the spectrum, and
hence allow in most cases to break the degeneracy between the hot and
cold solutions. Figure \ref{wdfits} illustrates this procedure. Once
\Teff\ and $\log g$ are determined, the white dwarf mass and radius
can be estimated using an updated version of
\citeauthor{bergeronetal95-2}'s (1995) tables. The results, after
applying this method to our four systems, are summarised in Table \ref{specres}.
The preferred solution (``hot'' or ``cold'') is highlighted in bold fond.

In the case of SDSS\,0303+0054, which contains a DC white dwarf
(Fig.\,\ref{spec}), the spectral decomposition results in 
$\mathrm{Sp}(2)=\mathrm{M}4.5\pm0.5$ for the secondary star. The
subsequent fit to the residual white dwarf spectrum is not physical
meaningful, as the white dwarf in SDSS\,0303+0054 does not exhibit
Balmer lines. The fit to the overall spectrum, which is purely
sensitive to the continuum slope, suggests $\Teff<8000$\,K, which is
consistent with the DC classification of the white dwarf. 

\begin{table}
\centering \setlength{\tabcolsep}{0.7ex}
\caption{Summary of the results obtained for SDSS\,0110+1326, SDSS\,1435+3733 and SDSS\,1548+4057 
from our spectral decomposition technique. The preferred set of parameters for each system is highlighted in bold fond.}
\label{specres}
\begin{tabular}{@{}ccccc@{}}
\hline
 Solution & \Mwd [\Msun] & $\log g$ & \Teff [$\mathrm{K}$]& $\mathrm{Sp}(2)$ \\
\hline
\multicolumn{5}{l}{\textbf{SDSS\,0110+1326}} \\ 
\textbf{Hot} & \textbf{0.47}$\pm$\textbf{0.02} & \textbf{7.65}$\pm$\textbf{0.05} & \textbf{25891}$\pm$\textbf{427} & \textbf{M4}$\pm$\textbf{1} \\
Cold & $1.2\pm0.03$ & $9\pm0.04$ & $9619\pm23$ & $\mathrm{M}4\pm1$ \\
\hline
\multicolumn{5}{l}{\textbf{SDSS\,1435+3733}} \\ 
Hot & $0.40\pm0.05$ & $7.58\pm0.11$ & $12536\pm438$ & $\mathrm{M}4.5\pm0.5$ \\
\textbf{Cold} & \textbf{0.41}$\pm$\textbf{0.05} & \textbf{7.62}$\pm$\textbf{0.12} & \textbf{12536}$\pm$\textbf{488} & \textbf{M4.5}$\pm$\textbf{0.5} \\
\hline
\multicolumn{5}{l}{\textbf{SDSS\,1548+4057}} \\ 
Hot & $0.43\pm0.16$ & $7.64\pm0.31$ & $14899\pm1300$ & $\mathrm{M}6\pm0.5$ \\
\textbf{Cold} & \textbf{0.62}$\pm$\textbf{0.28} & \textbf{8.02}$\pm$\textbf{0.44} & \textbf{11699}$\pm$\textbf{820} & \textbf{M6}$\pm$\textbf{0.5} \\
\hline
\end{tabular}
\end{table}

%-------------------------------------------------------------------------------------------------------%
%-------------------------------------------------------------------------------------------------------%

\section{Basic equations}
\label{s-equations}

Here, we introduce the set of equations that we will use to constrain
the stellar parameters of our four eclipsing WDMS binaries.  In a
binary system, where a white dwarf primary and a main-sequence  companion, 
with masses \Mwd\ and \Msec\ respectively, orbit
with a period $P_{\mathrm{orb}}$ around their common centre of mass at a
separation $a=a_\mathrm{WD}+a_\mathrm{sec}$ where $a_\mathrm{WD}\,\Mwd\,=\,a_\mathrm{sec}\,\Msec$, the orbital velocity $K$ of either star, as observed at an inclination angle $i$, is

\begin{equation}
\label{kvel}
 K_{j}=\frac{2\pi a_{j}}{P_{\mathrm{orb}}}\sin i \qquad j=\mathrm{WD, sec}
\end{equation}

\noindent
assuming circular orbits. Given Kepler's law

\begin{equation}
\label{kep}
 a^{3}=\frac{\mathrm{G}\left(\Mwd+\Msec\right)P^{2}_{\mathrm{orb}}}{4\pi^{2}}
\end{equation}
\noindent
and using

\begin{equation}
\label{aeq}
 a=a_\mathrm{WD}\frac{\Mwd+\Msec}{\Msec}
\end{equation}
\noindent
one can obtain the two mass functions:

\begin{equation}
\label{k1eq}
 f\left(\Msec\right)=\frac{\left(\Msec\sin i\right)^{3}}{\left(\Mwd+\Msec\right)^{2}}=\frac{P_{\mathrm{orb}}K^{3}_\mathrm{WD}}{2\pi G}\quad<\Msec
\end{equation}

\begin{equation}
\label{k2eq}
 f\left(\Mwd\right)=\frac{\left(\Mwd\sin i\right)^{3}}{\left(\Mwd+\Msec\right)^{2}}=\frac{P_{\mathrm{orb}}K^{3}_\mathrm{sec}}{2\pi G}\quad<\Mwd
\end{equation}

\noindent
which give strict lower limits on the masses of the
components. From these two equations we get

\begin{equation}
\label{qeq}
 q=\frac{\Msec}{\Mwd}=\frac{K_\mathrm{WD}}{K_\mathrm{sec}}
\end{equation}

\noindent
Therefore, the knowledge of the radial velocities of both stars can
immediately yield the mass ratio $q$ of the system, using
Eq.\,(\ref{qeq}). In our case, since we lack a measurement for the radial
velocity of the white dwarf $K_\mathrm{WD}$, a more indirect approach needs to
be followed. Re-arranging Eq.\,(\ref{k2eq}) for $\sin i$ yields

\begin{equation}
 \label{ieq}
\sin i=\left[\frac{P_{\mathrm{orb}}K^{3}_\mathrm{sec}}{2\pi
    G}\frac{\left(\Mwd+\Msec\right)^2}{M^{3}_{\mathrm{WD}}}\right]^{1/3}
\end{equation}

\noindent
whereas re-arranging Eq.\,(\ref{k2eq}) for $K_\mathrm{sec}$ yields

\begin{equation}
 \label{k2calc}
K_\mathrm{sec}=\left[\frac{2\pi G\sin^{3}i}{P_{\mathrm{orb}}}\frac{M^{3}_{\mathrm{WD}}}{\left(\Mwd+\Msec\right)^{2}}\right]^{1/3}
\end{equation}

We make use of Equations (\ref{ieq}) and (\ref{k2calc}) in the light
curve fitting process described in the next section.

\begin{figure}
 \includegraphics[width=80mm]{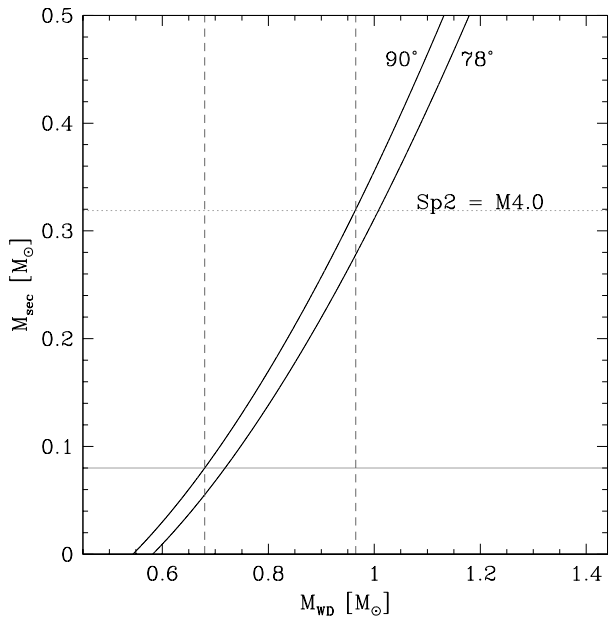}
  \caption{Mass function plot (see Eq. \ref{k2eq}) for
    SDSS\,0303+0054, for $i=78^{\circ}$ and $i=90^{\circ}$. The gray
    horizontal line corresponds to a lower mass limit for
    M-dwarfs. The dotted horizontal line is the value of \Msec,
    corresponding to a secondary spectral type of
    $\mathrm{Sp}(2)=\mathrm{M}4$, assuming a $\mathrm{Sp}(2)-\Msec$
    relation. The dashed vertical lines indicate the corresponding
    range for the white dwarf mass.}
  \label{mf0303} 
\end{figure}

For the DC white dwarf in SDSS\,0303+0054, where we lack a
spectroscopic estimate of \Mwd, we can use Eq.\,(\ref{k2eq}) to get a
rough, first estimate of the white dwarf mass. In Fig.\,\ref{mf0303}, 
we have plotted Eq.\,(\ref{k2eq}) for $i=90^{\circ}$ and
$i=78^{\circ}$. For lower inclinations no eclipses occur for
$K_\mathrm{sec}=339.7\,\kms$ and $\Porb=0.134$\,d. For a given choice
of the inclination angle, Eq.\,(\ref{k2eq}) defines a unique relation
$\Msec\left(\Mwd\right)$. If we further assume a mass for the
secondary, we can investigate the possible range of the white dwarf
mass for SDSS\,0303+0054. For an extremely conservative lower limit of
$\Msec\ge0.08\,\Msun$ (the lower mass limit for an M-dwarf, e.g.
\citet{dormanetal89-1}, with the spectrum of SDSS\,0303+0054 clearly
identifying the companion as a main-sequence star),
$\Mwd\ge0.68\,\Msun$. Even under this extreme assumption for the
companion star, the white dwarf in SDSS\,0303+0054 has to be more
massive than the average field white dwarf \citep{koesteretal79-1,
  liebertetal05-1}.  If we assume $\mathrm{Sp}(2)=\mathrm{M}4$, the
upper limit of the spectral type according to our spectral
decomposition of the SDSS spectrum, and use the spectral type-mass
relation of \citet{rebassa-mansergasetal07-1}, we find
$\Msec\simeq0.32\,\Msun$, and hence
$\Mwd\ge0.96\,\Msun$. These estimates assume $i=90^{\circ}$, for
lower inclinations the respective \Mwd\ values become larger, as
illustrated in Fig.\,\ref{mf0303}. In summary, based on
$K_\mathrm{sec}$, $\Porb$, and a generous range in possible \Msec\, we
expect the white dwarf in SDSS\,0303+0054 to be fairly massive,
$0.7\Msun\la\Mwd\la1.0\Msun$.

%-------------------------------------------------------------------------------------------------------%
%-------------------------------------------------------------------------------------------------------%

\section{Light Curve model fitting}
\label{s-lcmodels}

Analysing the light curves of eclipsing binaries can provide strong
constraints on the physical parameters of both stars.  Here, we make
use of a newly developed code written by one of us (TRM) for the
general case of binaries containing a white dwarf. The code offers the
option to include accretion components (disc, bright spot) for the
analysis of cataclysmic variables~--~given the detached nature of our
targets, those components were not included. The program sub-divides
each star into small elements with a geometry fixed by its radius as
measured along the direction of centres towards the other star. The
code allows for the distortion of Roche geometry and for irradiation
of the main-sequence star using the approximation $\sigma
T_\mathrm{sec}^{'4} = \sigma T_\mathrm{sec}^{4} + F_\mathrm{irr}$,
where $T'_\mathrm{sec}$ is the modified temperature and $T_\mathrm{sec}$ the
temperature of the unirradiated companion, $\sigma$ is the
Stefan-Boltzmann constant and $F_\mathrm{irr}$ is the irradiating flux,
accounting for the angle of incidence and distance from the white
dwarf. The white dwarf is treated as a point source in this
calculation and no backwarming of the white dwarf is included. The
latter is invariably negligible, while the former is an unnecessary
refinement given the approximation inherent in treating the
irradiation in this simple manner.

The code computes a model based on input system parameters supplied by
the user. Starting from this parameter set, model light curves are
then fitted to the data using Levenberg-Marquardt minimisation, where
the user has full flexibility as to which parameters will be optimised
by the fit, and which ones will be kept fixed at the initial value. 

The physical parameters, which define the models, are the radii scaled
by the binary separation, $\Rwd/a$ and $\Rsec/a$, the orbital
inclination, $i$, the unirradiated stellar temperatures of the white dwarf
and the secondary star \Twd\ and \Tsec\ respectively, 
the mass ratio $q = \Msec/\Mwd$, $T_0$ the time of mid-eclipse of
the white dwarf and $d$ the distance. We account for the distance
simply as a scaling factor that can be calculated very rapidly for any
given model, and so it does not enter the Levenberg-Marquardt
optimisation.  All other parameters can be adjusted, i.e. be
allowed to vary during the fit, but typically the light-curve of a
given system does not contain enough information to constrain all of
them simultaneously. For instance, for systems with negligible
irradiation, fitting \Twd, \Tsec\ and $d$ simultaneously is degenerate
since a change in distance can be exactly compensated by changes in
the temperatures.

Some of our exposures were of a significant length compared to the
length of the white dwarf's ingress and egress and therefore we
sub-divided the exposures during the model calculation, trapezoidally
averaging to obtain the estimated flux.

Our aim is to combine the information from the $K_\mathrm{sec}$ radial
velocity amplitudes determined in Sect.\,\ref{s-k2}, the information
contained in the light curve, and the white dwarf effective
temperature determined from the spectral fit in
Sect.\,\ref{s-spectralfit} to establish a full set of stellar
parameters for the four eclipsing WDMS binaries. The adopted method is
described in the next two subsections, where SDSS\,0110+1326 requires
a slightly broader approach, as a correction needs to be applied to
the $K_\mathrm{sec}$ velocity determined from the emission lines.
 
\subsection{SDSS\,0303+0054, SDSS\,1435+3733 and SDSS\,1548+4057}
\label{s-lcmodels0303etal}

Our approach is to fit a light curve model to the data for a selected
grid of points in the $\Mwd-\Msec$ plane. Each point in the
$\Mwd-\Msec$ plane defines a mass ratio $q$ (Eq.\,\ref{qeq}), and,
using $K_\mathrm{sec}$ and $\Porb$, a binary inclination $i$
(Eq.\,\ref{ieq}). Hence, a light curve fit for a given (\Mwd, \Msec)
combination will have \Mwd, \Msec, $q$, and $i$ fixed, whereas \Rwd,
\Rsec, \Tsec, \Twd, and $T_0$ are in principle free parameters.  In
practice, we fix \Twd\ to the value determined from the spectroscopic
fit, which is sufficiently accurate, hence only \Rwd, \Rsec, \Tsec,
and $T_0$ are free parameters in the light curve fits. We leave $T_{0}$
free to vary to account for the $O-C$ errors in each individual light curve.
The fitted values of $T_{0}$ were of the order of 10\,s, consistent 
with the $O-C$ values quoted in Table\,\ref{midtimes}. Each light curve fit in the
$\Mwd-\Msec$ plane requires some initial estimates of \Rwd, \Rsec,
\Tsec. For \Rwd\, we adopt the theoretical white dwarf mass-radius
relation interpolated from \citeauthor{bergeronetal95-2}'s (1995)
tables. For \Rsec\, we use the mass-radius relation of
\citet{baraffeetal98-1}, adopting an age of 5\,Gyr. For \Tsec, we use
the spectral type of the secondary star as determined from the
spectral decomposition (Sect.\,\ref{s-spectralfit}) combined with the
spectral type-temperature relation from
\citet{rebassa-mansergasetal07-1}. \Rwd\ and \Rsec\ are then scaled by
the binary separation, Eq.\,(\ref{aeq}).

We defined large and densely covered grids of points in the
$\Mwd-\Msec$ plane which generously bracket the initial estimates for
\Mwd\ and \Msec\ based on the spectral decomposition/fitting. For
SDSS\,0303+0054, where no spectral fit to the white dwarf is
available, we bracket the range in \Mwd\ illustrated in
Fig.\,\ref{mf0303}. Points for which (formally) $\sin i > 1$ were
discarded from the grid, for all other points a light curve fit was
carried out, recording the resulting \Rwd, \Rsec, and
\Tsec. The number of light curve fits performed for each system was 
between 7000 and 10000, depending on the system. 

\subsection{SDSS\,0110+1326}
\label{s-lcmodels0110}

In the case of SDSS\,0110+1326 the inclination angle cannot readily be
calculated through Eq.\,(\ref{ieq}), as we measured $K_\mathrm{sec}$
from either the H$\alpha$ or \Ion{Ca}{II} emission lines, which do not
trace the motion of the secondary's centre of mass, but the centre of
light in the given emission line. The general approach in this case is to apply a correction to
the observed $K_\mathrm{sec}$, where the motion of the secondary
star's centre of mass, $K_{\mathrm{sec\,,cor}}$ is expressed according
to \citet{wade+horne88-1} as

\begin{equation}
 \label{k2corr}
K_{\mathrm{sec,\,cor}}=\frac{K_\mathrm{sec}}{1-\left(1+q\right)\left(\Delta
  R/a\right)}
\end{equation}
where $K_\mathrm{sec}$ is the measured radial velocity, $q$ is
the mass ratio of the system, $a$ the binary separation and $\Delta$R
is the displacement of the centre of light from the centre of mass of
the secondary. $\Delta R$ can have a minimum value of zero, i.e. the
two centres coincide and no correction is needed and a maximum value
of $R_{\mathrm{sec}}$, i.e. all light comes from a small region of the
secondary star closest to the primary. An assumption often used in the
literature is that the emission due to irradiation is distributed
uniformly over the inner hemisphere of the secondary star, and is zero
on its unirradiated side, in which case $\Delta R =
\left(4/3\pi\right)$\Rsec\ \citep[e.g.][]{wade+horne88-1,
  woodetal95-3, oroszetal99-1, vennesetal99-2}. A more physical model
of the distribution of the irradiation-induced emission line flux can
be derived from the analysis of the orbital variation of the
equivalent width of the emission line. A fine example of this approach
is the detailed study of the \Lines{Ca}{II}{8498.02, 8542.09,8662.14}
emission in the WDMS binary HR\,Cam presented by
\citet{maxtedetal98-1}. However, given the small number of spectra, it
is currently not possible to apply this method to SDSS\,0110+1326.

\begin{figure}
 \includegraphics[width=84mm]{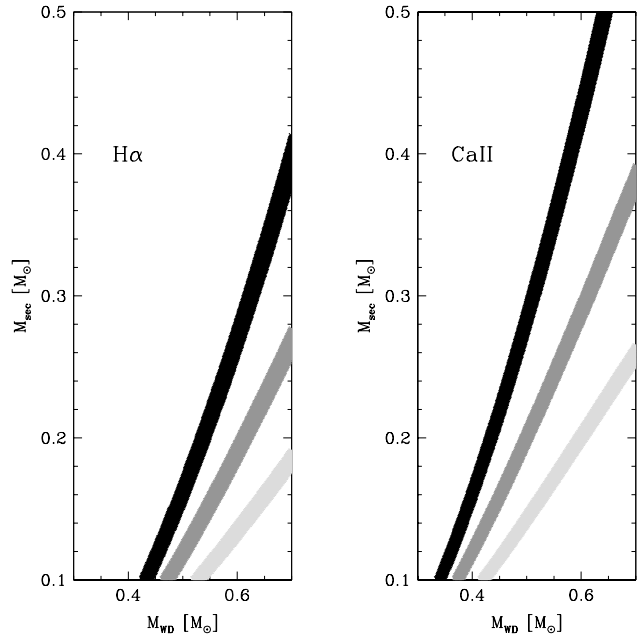}
  \caption{Grid of $(\Mwd,\Msec)$ points for which the
    light curve of SDSS\,0110+1326 has been fitted.  Different
    assumptions for the $K$-correction were made: $\Delta R=0$ (black
    strip), $\Delta R=(4/3\pi)\Rsec$ (dark gray strip) and
    $\Delta R=\Rsec$ (light gray strip). Left panel: (\Mwd,\Msec) grid
    for $K_\mathrm{sec}=K_{\mathrm{sec,\,H}\alpha}=200.1\,\kms$. Right panel:
    (\Mwd,\Msec) grid for $K_\mathrm{sec}=K_{\mathrm{sec,\,\Ion{Ca}{II}}}=178.8\,\kms$.}
  \label{k2cases} 
\end{figure}

Our approach for modelling SDSS\,0110+1326 was to calculate the binary
separation from Eq.\,(\ref{kep}) for each pair of
$\left(\Mwd,\Msec\right)$ of the initial grid and then, using
Eq.\,(\ref{k2corr}), calculate various possible $K_{\mathrm{sec,\,cor}}$
values for this $\left(\Mwd,\Msec\right)$ point, assuming various
forms of $\Delta R$. We adopted the following three different cases:
(i)~$\Delta R=0$, i.e. the centre of light coincides with the centre
of mass, (ii)~$\Delta R = \left(4/3\pi\right)$\Rsec, i.e. the uniform
distribution case and (iii)~$\Delta R=$\Rsec, i.e. the maximum
possible displacement of the centre of light. This was done for both
the $\mathrm{H}\alpha$ and the \Ion{Ca}{II} lines. Having attributed a
set of corrected radial velocity values to each point of the
grid, we could now make use of Eq.\,(\ref{ieq}) to calculate the
corresponding inclination angles. Again, points with a (formal) $\sin
i>1$ were eliminated from the grid. The allowed
$\left(\Mwd,\Msec\right)$ pairs for both lines and all three $\Delta R$
cases are shown in Fig.\,\ref{k2cases}.

The first of the three $\Delta R$ cases, $\Delta R=0$, is obviously
not a physically correct approach, as it implies that no correction is
necessary. However, we did use it as a strict lower limit of the
radial velocity of the secondary. In this sense, the third case,
$\Delta R=$\Rsec, is a strict upper limit for
$K_\mathrm{sec}$. Table\,\ref{rvallowed} lists the ranges
$K_{\mathrm{sec},\mathrm{cor}}$ corresponding the our adopted grid in
the $\left(\Mwd,\Msec\right)$ plane shown in Fig.\,\ref{k2cases}.

\begin{table}
\centering \setlength{\tabcolsep}{0.8ex}
\caption{Corrected values of the radial velocity of the
  secondary star in SDSS\,0110+1326, after assuming a form of $\Delta
  R$, for both the cases of the $\mathrm{H}\alpha$ and the
  \Ion{Ca}{II} emission lines.} 
\label{rvallowed}
\begin{tabular}{@{}cccc@{}}
\hline 
$\Delta R$ & mean [\kms] & std.dev. & range [\kms] \\
\hline
\multicolumn{4}{l}{{$\mathbf{\mathrm{H}\alpha}$}} \\
0 & 200.1 & - & - \\
$\left(4/3\pi\right)R_{\mathrm{sec}}$ & 213.1 & 2.7 & 207.9-218.2 \\
$R_{\mathrm{sec}}$ & 225.2 & 3.5 & 218.5-232.1 \\
\hline 
\multicolumn{4}{l}{\textbf{\Ion{Ca}{II}}} \\
0 & 178.8 & - & - \\ 
$\left(4/3\pi\right)R_{\mathrm{sec}}$ & 195.1 & 4.2 & 186.6-202.7 \\
$R_{\mathrm{sec}}$ & 209.1 & 6.1 & 197.2-219.9 \\
\hline
\end{tabular}
\end{table}

%-------------------------------------------------------------------------------------------------------%
%-------------------------------------------------------------------------------------------------------%

\section{Results}
\label{s-results}

Our light curve fitting procedure outlined in Sect.\,\ref{s-lcmodels}
yields fitted values for \Rwd, \Rsec, and \Tsec\ for a large grid in
the $\Mwd-\Msec$ plane, where each $(\Mwd,\Msec)$ point defines $q$
and $i$. Analysing these results was done in the following fashion.

In a first step, we applied a cut in the quality of the light curve
fits, where we considered three different degrees of ``strictness'' by
culling all models whose $\chi^2$ was more than 1, 2 or 3$\sigma$
above the best-fit value. This typically left us with a relatively
large degeneracy in \Mwd\ and \Msec.

In a second step, we had to select among various equally good light
curve fits those that are physically plausible. Given that our set of
fitted parameters was underconstrained by the number of observables,
we had to seek input from theory in the form of mass-radius relations
for the white dwarf (using the tables of \citealt{bergeronetal95-2})
and for the M-dwarf (using the $5.0$\,Gyr model of
\citealt{baraffeetal98-1}). We then calculated for each model the
relative difference between the theoretical value of the radius and
the value of the radius from the fit,
\begin{equation}
\delta R =
\left|\frac{R_{\mathrm{fit}}-R_{\mathrm{th}}}{R_{\mathrm{th}}}\right|,
\end{equation}
where $R_\mathrm{fit}$ and $R_\mathrm{th}$ are the radius obtained
from the light curve fit, and the radius from the mass-radius
relation, respectively. Assuming that the binary components obey, at
least to some extent, a theoretical mass-radius relation, we defined
cut-off values of $\delta R=5\%$, 10\%, and 15\%, above which light
curve models will be culled. The maximum of $\delta R=15\%$ was
motivated by the radius excess over theoretical main-sequence models
observed in eclipsing low-mass binaries \citep{ribasetal07-1}, and by
the current observational constraints on the white dwarf mass-radius
relation \citep{provencaletal98-1}.  This cut in radius was applied
individually for the white dwarf, ($\delta R_{\mathrm{WD}}$) and the
companion star ($\delta R_{\mathrm{sec}}$).

Combining both the $\chi^2$ and $\delta R$ cuts left us in general with
a relatively narrow range of system parameters that simultaneously
satisfy the radial velocity amplitude and the morphology of the light
curve. In a final step, we examined how well the stellar masses
determined from  the light curve/radial velocity analysis agreed with
those derived from the spectral decomposition. 

\subsection{SDSS\,0303+0054}

Modelling the light curve of SDSS\,0303+0054 involved the following
two problems. Firstly, the time resolution of our data set poorly
resolves the white dwarf ingress and egress phases,
and consequently the white dwarf radius can only be loosely
constrained. We decided therefore to fix the white dwarf radius in the
light curve fits to the value calculated from the adopted white dwarf
mass-radius relation of \citealt{bergeronetal95-2}. Secondly, the light curve displays a strong, but
slightly asymmetric ellipsoidal modulation, with the two maxima being
of unequal brightness (Fig.\,\ref{all}). Similar light curve
morphologies have been observed in the WDMS binaries BPM\,71214
\citep{kawka+vennes03-1} and LTT\,560 \citep{tappertetal07-1}, and
have been attributed to star spots on the secondary star. By
design, our light curve model can not provide a fit that
reproduces the observed asymmetry. Having said this, the presence of
ellipsoidal modulation provides an additional constraint on \Rsec\ that
is exploited by the light curve fit.

\begin{figure}
 \includegraphics[width=84mm]{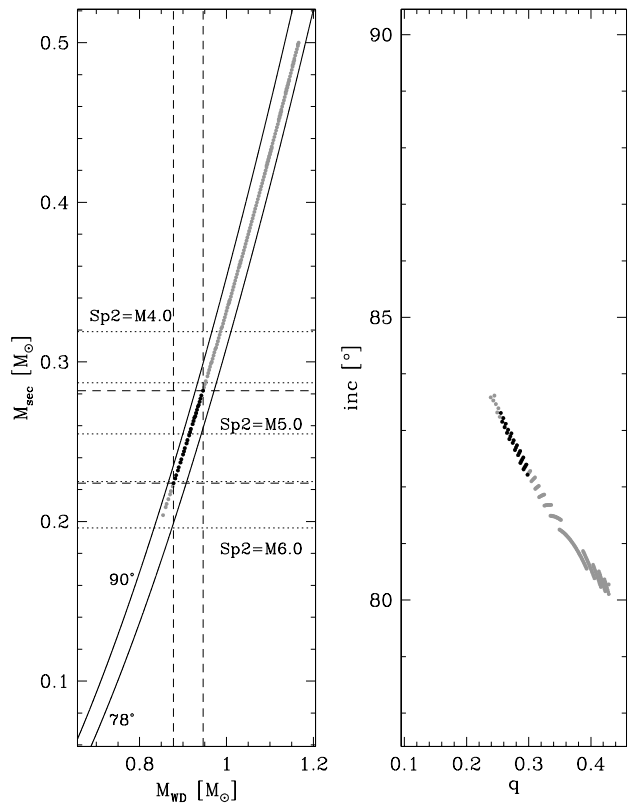}
  \caption{Light curve model fitting results for SDSS\,0303+0054. Left
    panel: \Mwd\, and \Msec\, values corresponding to fits with $\chi^{2}$
    values within $1\sigma$ of the minimum value (gray points) and,
    simultaneously, with $\delta\Rsec\le0.05$ (black points). Right
    panel: the same, only in the $q-i$ plane. Also depicted in the
    left panel are curves corresponding to the mass function (solid
    black lines, $i=90^{\circ}$ and $78^{\circ}$) which (by
    definition) bracket the possible solutions, $\mathrm{Sp}(2)-\Msec$
    relations (dotted, horizontal, black lines) and the range of possible
    $\left(\Mwd,\Msec\right)$ values (dashed, horizontal and vertical,
    black lines)}
  \label{0303fitall} 
\end{figure}

A number of models passed the strictest configuration of our cut-offs,
i.e. $1\sigma$ and $\delta \Rsec=5\%$ (no cut was used in
$\delta \Rwd$, as \Rwd\ has been fixed during the fits). The possible
range of solutions in the $\Mwd-\Msec$ plane is shown in
Fig.\,\ref{0303fitall}.

\begin{figure}
 \includegraphics[angle=-90,width=84mm]{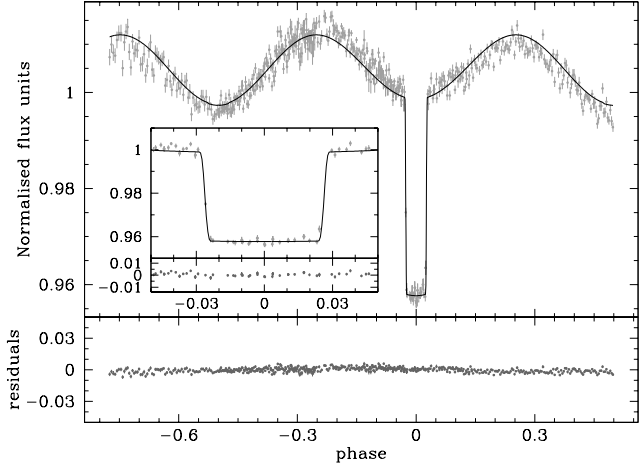}
  \caption{Model fit to the CA2.2 filterless light curve of SDSS\,0303+0054, for
    $\Mwd=0.91\,\Msun$, $\Rwd=0.009\,\Rsun$, $\Msec=0.25\,\Msun$,
    $\Rsec=0.26\,\Rsun$ and $i=82.6^{\circ}$. The
    model meets both our $\chi^{2}$ (within $1\sigma$) and the $\delta
    R$ (within $5\%$) cut-offs. The residuals from the fit are shown
    at the bottom of the panel. Inset panel: data points and model fit
    focused around the eclipse.}
  \label{0303fit} 
\end{figure}

Gray dots designate those light curve fits making the 1$\sigma$ cut,
black dots those that satisfy both the 1$\sigma$ and $\delta \Rsec=5\%$
cuts.  The resulting ranges in white dwarf masses and secondary star
masses are $\Mwdl=0.88-0.95\,\Msun$ and $\Msec=0.22-0.28\,\Msun$,
respectively, corresponding to a white dwarf radius of
$\Rwd=0.008-0.009\,\Rsun$ and a secondary radius of
$\Rsec=0.25-0.27\,\Rsun$. Fig.\,\ref{0303fit} shows one example of the
light curve fits within this range for the model parameters
$\Mwd=0.91\,\Msun$, $\Rwd=0.009\,\Rsun$, $\Msec=0.25\,\Msun$, 
$\Rsec=0.26\,\Rsun$ and $i=82.6^{\circ}$.

To check the consistency of our light curve fits with the results from
the spectral decomposition, we indicate in Fig.\,\ref{0303fitall} the
radii of M-dwarfs with spectral types
$\mathrm{Sp}(2)=\mathrm{M}4-\mathrm{M}6$ in steps of 0.5, based on the
spectral type-mass relation given by
\citet{rebassa-mansergasetal07-1}. The secondary star mass from the
light curve fit, $\Msecl=0.23-0.28\,\Msun$, corresponds to a
secondary spectral type of
$\mathrm{Sp}(2)_\mathrm{lcfit}=\mathrm{M}4.5-\mathrm{M}5.5$. The expected spectral type
of the secondary from the spectral decomposition
(Sect.\,\ref{s-spectralfit}) was 
$\mathrm{Sp}_\mathrm{spfit}(2)=\mathrm{M}4.5\pm0.5$, hence, both methods yield
consistent results. 

\subsection{SDSS\,1435+3733}

For SDSS\,1435+3733, which is partially eclipsing, our temporal
resolution, although again not ideal, was nevertheless deemed to be
adequate to constrain the white dwarf radius from the light
curve fits. Thus, a $\delta \Rwd$ cut was also applied.

\begin{figure}
 \includegraphics[width=84mm]{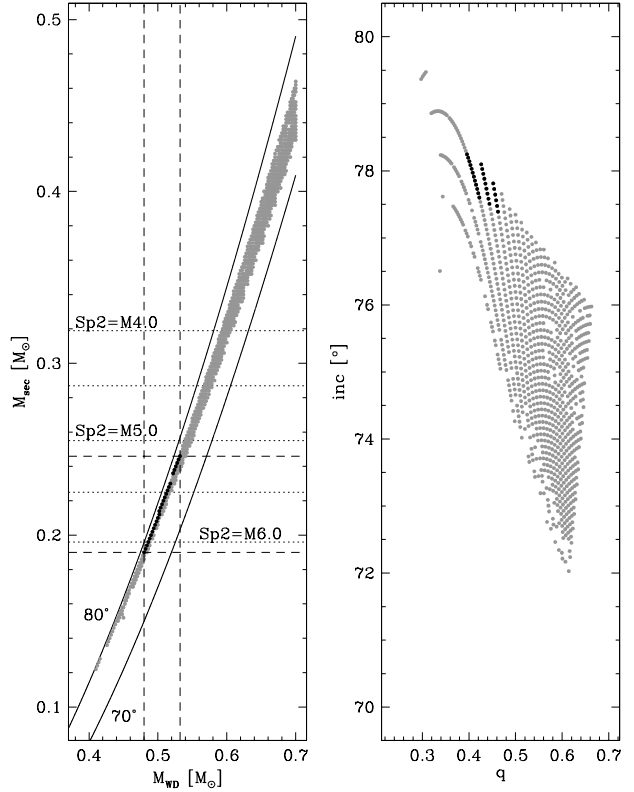}
  \caption{Light curve model fitting results for SDSS\,1435+3733. Left
    panel: \Mwd\, and \Msec\, values corresponding to fits with $\chi^{2}$
    values within $1\sigma$ of the minimum value (gray points) and,
    simultaneously for both white dwarf and secondary radii, with
    $\delta R\le0.05$ (black points). Right panel: the same, only in the
    $q-i$ plane. Also depicted in the left panel are mass functions
    ($i=70^{\circ}$ and $i=80^{\circ}$, solid black lines),
    $\mathrm{Sp}(2)-\Msec$ relations (dotted, horizontal, black lines) 
    and the range of possible
    $\left(\Mwd,\Msec\right)$ values (dashed, horizontal and vertical,
    black lines)}
  \label{1435fitall} 
\end{figure}

We applied again our strictest cuts on the light curve models,
$1\sigma$ and $\delta \Rwd=\delta \Rsec=5\%$, and the parameters of
the surviving models are shown in Fig.\,\ref{1435fitall}, where the
meaning of the symbols is the same as in  Fig.\,\ref{0303fitall}. 

\begin{figure}
 \includegraphics[angle=-90,width=84mm]{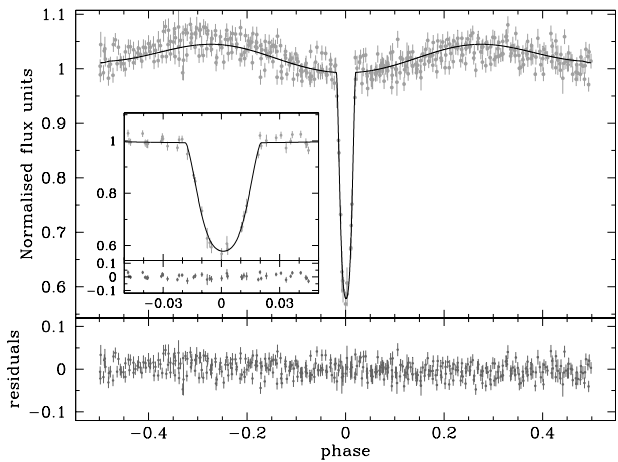}
  \caption{Model fit to IAC80 $I$-band light curve of SDSS\,1435+3733, for
    $\Mwd=0.5\,\Msun$, $\Rwd=0.015\,\Rsun$, $\Msec=0.21\,\Msun$,
    $\Rsec=0.23\,\Rsun$ and $i=77.6^{\circ}$.  The
    model meets both the $1\sigma$ cut in $\chi^{2}$, and a $\delta
    R\le0.05$ cut for both the white dwarf and the secondary star.  The
    residuals from the fit are shown at the bottom of the panel. Inset
    panel: data points and model fit focused around the eclipse.}
  \label{1435fit} 
\end{figure}

These fits imply a white dwarf mass and radius of
$\Mwd=0.48-0.53\,\Msun$ and $\Rwd=0.014-0.015\,\Rsun$, respectively,
and a secondary star mass and radius of $\Msec=0.19-0.25\,\Msun$ and
\Rsec$=0.22-0.25\,\Rsun$, respectively. A sample fit that obeyed all
three constraints is shown in Fig. \ref{1435fit}. The model parameters
are $\Mwd=0.5\,\Msun$, $\Rwd=0.015\,\Rsun$, $\Msec=0.21\,\Msun$, \Rsec$=0.23\,\Rsun$ and $i=77.6^{\circ}$.

Comparing the white dwarf masses from the spectroscopic
decomposition/fit, $\Mwds=0.41\pm0.08\,\Msun$, with the range of white
dwarf masses allowed by the light curve fitting,
$\Mwdl=0.48-0.53\,\Msun$, reveals a reasonable agreement. Regarding
the spectral type of the secondary star, we indicate in
Fig.\,\ref{1435fitall} again the masses of M-dwarfs in the range
M4--M6, in steps of 0.5 spectral classes, following the
$\mathrm{Sp}(2)-\Msec$ relation of
\citet{rebassa-mansergasetal07-1}. This illustrates that the light
curve fits result in a
$\mathrm{Sp}(2)_\mathrm{lcfit}=\mathrm{M}5-\mathrm{M}6$, whereas the
spectral decomposition provided
$\mathrm{Sp}(2)_\mathrm{spfit}=\mathrm{M}4.5\pm0.5$. Hence, also the
results for the secondary star appear broadly consistent.

\begin{figure}
 \includegraphics[angle=-90,width=84mm]{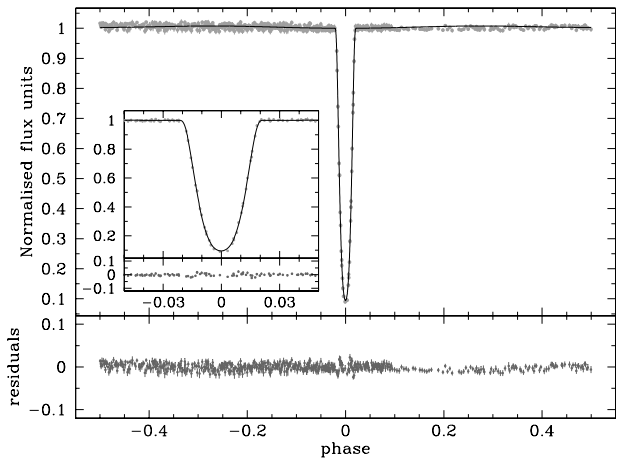}
  \caption{Our light curve model for SDSS\,1435+3733, as determined
    from fitting our IAC80 $I$-band photometry as shown in
    Fig.\,\ref{1435fit}, along with the data from
    \citet{steinfadtetal08-1}, illustrating a high degree of consistency.}
  \label{1435just} 
\end{figure}

\citet{steinfadtetal08-1} first reported the eclipsing nature of
SDSS\,1435+3733, and estimated $\Mwd=0.35-0.58\,\Msun$,
$\Rwd=0.0132-0.0178\,\Rsun$, $\Msec=0.15-0.35\,\Msun$ and
$\Rsec=0.17-0.32\,\Rsun$. The parameter ranges determined from our
analysis ($\Mwd=0.48-0.53\,\Msun$, $\Rwd=0.014-0.015\,\Rsun$, $\Msec=0.19-0.25\,\Msun$ and
\Rsec$=0.22-0.25\,\Rsun$) are fully consistent with
\citeauthor{steinfadtetal08-1}'s work. In
Fig.\,\ref{1435just} we overplot on the photometry of
\citet{steinfadtetal08-1} the light curve model shown in
Fig.\,\ref{1435fit} along with our data, illustrating that our
solution is consistent with their data. 

While the temporal
resolution of our photometry is worse than that of
\citet{steinfadtetal08-1}, our analysis benefitted from two additional
constraints, firstly the mass function (Eq.\,\ref{k2eq}) determined
from our spectroscopy, and secondly the detection of a weak ellipsoidal
modulation in our $I$-band light curve (Fig.\,\ref{1435fit}). While the
observations of \citet{steinfadtetal08-1} covered the entire binary
orbit, their data were obtained through a BG-39 filter, centred at
4800\,\AA, where the flux contribution of the companion star is
negligible. A well-established difficulty in modelling light curves of
partially eclipsing binary stars is the fact that, although the sum
of the two radii can be accurately defined, their ratio can only be
loosely constrained \citep[e.g.][]{southworthetal07-3}. \Rwd\ and
\Rsec\ are strongly correlated, and fitting both of them leads to
degeneracy. Additional information is needed to lift this degeneracy,
which is mainly provided by ellipsoidal modulation or reflection effects
\citep[e.g.][]{hilditchetal96-2, hilditchetal96-1}.

\subsection{SDSS\,1548+4057}

\begin{figure}
 \includegraphics[width=84mm]{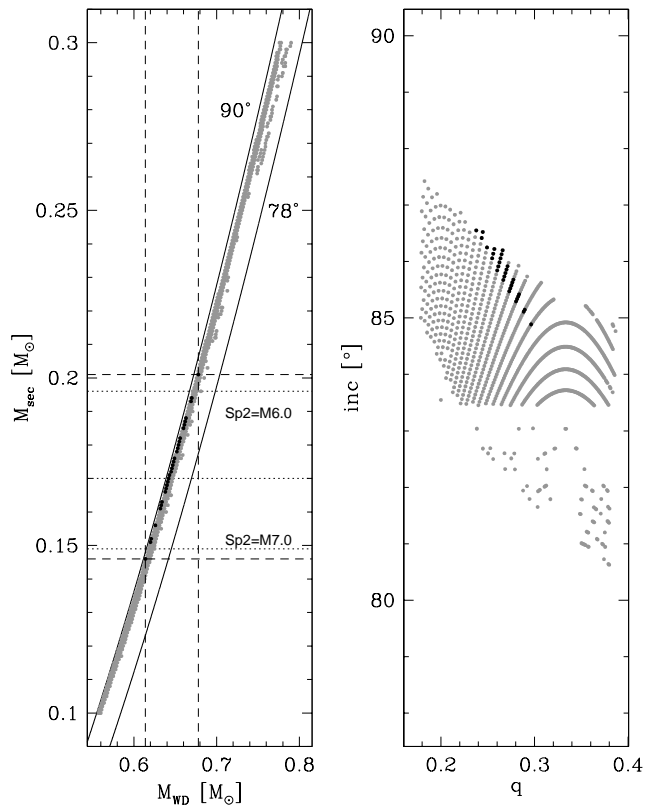}
  \caption{Light curve model fitting results for SDSS\,1548+4057. Left
    panel: \Mwd\, and \Msec\, values corresponding to fits with $\chi^{2}$
    values within $1\sigma$ of the minimum value (gray points) and,
    simultaneously for both white dwarf and secondary radii, with
    $\delta R\le0.1$ (black points). Right panel: the same, only in the
    $q-i$ plane. Also depicted in the left panel are mass function for
    $i=78^{\circ}$ and $i=90^{\circ}$ (solid black lines),
    $\mathrm{Sp}(2)-\Msec$ relations for spectral types M6, 6.5, and
    7 (dotted, horizontal, black lines) and the range of possible
    $(\Mwd,\Msec)$ values (dashed, horizontal and vertical, black
    lines).}
  \label{1548fitall} 
\end{figure}

\begin{figure}
 \includegraphics[angle=-90,width=84mm]{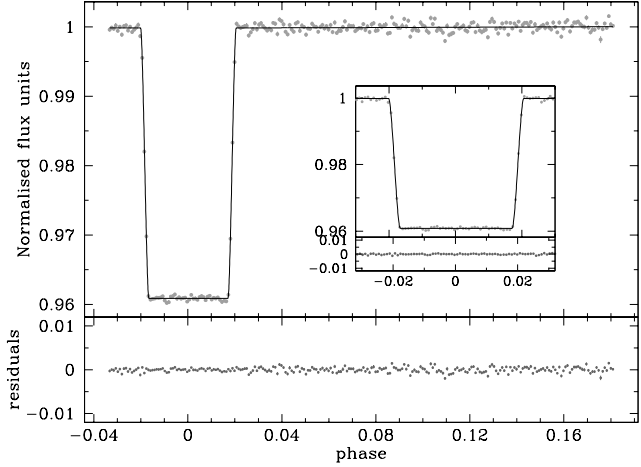}
  \caption{Model fit to the  WHT/$R$-band light curve of SDSS\,1548+4057, for
    $\Mwd=0.65\,\Msun$, $\Rwd=0.011\,\Rsun$, $\Msec=0.18\,\Msun$,
    $\Rsec=0.183\,\Rsun$ and $i=86.6^{\circ}$. The
    model meets both a $1\sigma$ $\chi^{2}$ cut-off, and the $\delta
    R\le0.10$, cut-off, for both the white dwarf and the secondary
    radii. The residuals from the fit are shown at the bottom of the
    panel. Inset panel: data points and model fit focused around the
    eclipse.}
  \label{1548fit} 
\end{figure}

For SDSS\,1548+4057, we follow a very similar approach as for
SDSS\,1435+3733, adopting a $1\sigma$ cut in $\chi^2$ but a slightly
less strict cut on the radii, $\delta\Rwd=\delta\Rsec=10\%$. 
Figure\,\ref{1548fitall} shows the solutions of our light curve fits
that survived those criteria, where plot symbols have the same meaning
as in Figs.\ref{0303fitall} and \ref{1435fitall}.

The models satisfying both cuts imply a white dwarf mass and radius of
$\Mwd=0.61-0.68\,\Msun$ and $\Rwd=0.010-0.011\,\Rsun$, respectively,
and a secondary mass and radius of $\Msec=0.15-0.20\,\Msun$ and
$\Rsec=0.17-0.20\,\Rsun$, respectively. A sample fit, that obeyed all
three constraints, is shown in Fig. \ref{1548fit}. The model
parameters are $\Mwd=0.65\,\Msun$, $\Rwd=0.011\,\Rsun$,
$\Msec=0.18\,\Msun$, $\Rsec=0.18\,\Rsun$
and $i=85.6^{\circ}$.

The white dwarf parameters obtained from the light curve fitting are
in full agreement with those derived from the spectral decomposition.
The former give a possible range of white dwarf masses of
$\Mwdl=0.61-0.68\,\Msun$, while the latter yield a value of
$\Mwds=0.62\pm0.28\,\Msun$. In Fig.\,\ref{1548fitall}, we indicate the
masses corresponding to secondary spectral types M6, M6.5, and M7. The
spectral fitting points to a
$\mathrm{Sp}(2)_{\mathrm{spfit}}=\mathrm{M}6$ secondary. Our models
indicate $\mathrm{Sp}(2)_{\mathrm{lcfit}}=\mathrm{M}6-\mathrm{M}7$
again consistent with the spectroscopic result.

\subsection{SDSS\,0110+1326}
Because of the uncertainty in the true $K_\mathrm{sec}$ velocity of
the secondary star, we inspected a total of six different grids of
light curve fits for SDSS\,0110+1326 (three K-corrections for each of
the observed $K_\mathrm{sec,\,H\alpha}$ and
$K_\mathrm{sec,\,\Ion{Ca}{II}}$, see Sect.\,\ref{s-lcmodels0110} and
Table\,\ref{results}), and consequently, this star needs a slightly
more extensive discussion compared to the other three systems.

\begin{figure}
 \includegraphics[width=84mm]{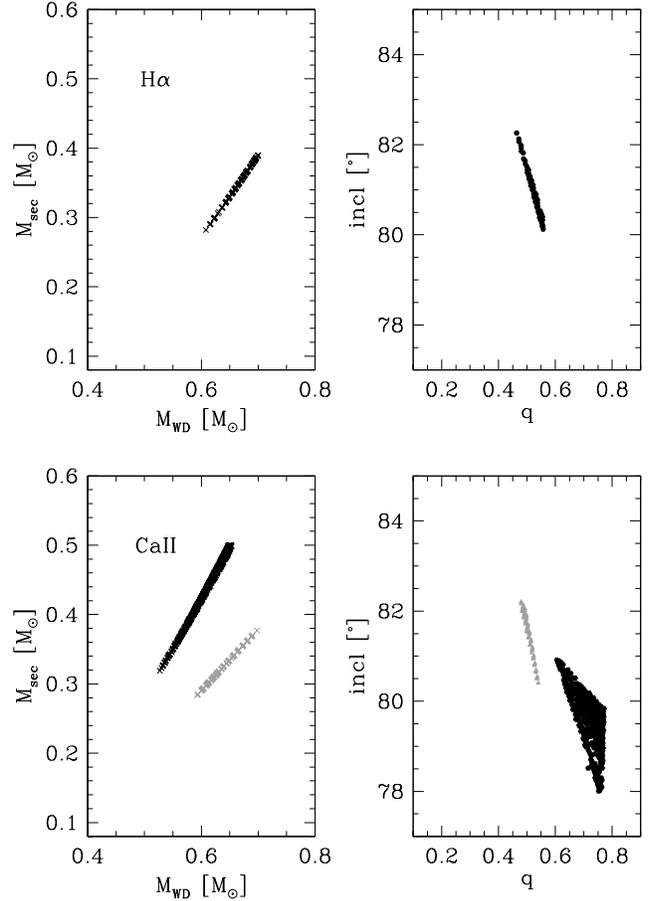}
  \caption{Results of the model fitting of SDSS\,0110+1326 for
    $\mathrm{H}\alpha$ (left panels) and \Ion{Ca}{II} (right panels),
    for a $3\sigma$ confidence level and $\delta R<0.15$. Black points
    correspond to $\Delta R=0$, gray to $\Delta
    R=\left(4/3\pi\right)$\Rsec (where applicable, see text for
    details). Top panels show the $\left(\Mwd,\Msec\right)$ plain,
    bottom panels the $q-i$ plain.}
  \label{0110fitresall} 
\end{figure}

In a first exploratory step, we applied our least strict cut-offs,
adopting a $3\sigma$ cut on the fit quality, and a $\delta R=15\%$ cut
for both the white dwarf and the secondary star radii. Even with these
loose constraints, the $\Delta R=\Rsec$ case, i.e. assuming that the
emission is concentrated on the surface of the secondary closest to
the white dwarf, does not produce any solutions that meet the cut-off
criteria. For the uniform illumination case, $\Delta
R=\left(4/3\pi\right)\Rsec$, there were no solutions for
$K_\mathrm{sec}=K_{\mathrm{sec,\,H}\alpha}$, but a few solutions for
the $K_\mathrm{sec}=K_{\mathrm{sec,\,\Ion{Ca}{II}}}$. Finally, a
number of solutions existed for the case $\Delta R=0$, i.e. assuming
no K-correction, for both H$\alpha$ and
\Ion{Ca}{II}. Table\,\ref{results} summarises this analysis, and the
models that survived the culling defined  ranges of (\Mwd,\Msec) pairs
which are illustrated in Fig.\,\ref{0110fitresall}.

\begin{table}
\centering 
\caption{Summary of the fitting results of SDSS\,0110+1326, for all
  three $\Delta R$ prescriptions, for both the cases of the
  $\mathrm{H}\alpha$ and the \Ion{Ca}{II} emission lines. In all cases
  we applied 3$\sigma$ and $\delta R = 15\%$ cut-offs. A ($\surd$)
  means some models passed the cut-off, whereas a (-) means no models
  passed.}
\label{results}
\begin{tabular}{@{}ccc@{}}
\hline
 & \textbf{$\mathbf{\mathrm{H}\alpha}$} & \textbf{\Ion{Ca}{II}} \\
\hline 
$\Delta R=0$ & $\surd$ & $\surd$ \\
\hline
$\Delta R=\left(4/3\pi\right)\Rsec$ & - & $\surd$ \\
\hline
$\Delta R=\Rsec$ & - & - \\
\hline
\end{tabular}
\end{table}

Given that the uniform irradiation prescription, $\Delta
R=\left(4/3\pi\right)\Rsec$ represents a mid-way between the two
extreme assumptions $\Delta R=\Rsec$ and $\Delta R=0$, and that a
reasonably large range of possible solutions was found for the
\Ion{Ca}{II} velocities (Fig.~\ref{0110fitresall}), we decided to
explore tighter $\chi^2$ constraints for this case. While a number of
light curve models are found within $1\sigma$ of the minimum
$\chi^{2}$ value, both $\delta\Rwd$ and $\delta\Rsec$ were $\sim15\%$ in
all cases. No models survived if we reduced the value of $\delta R$
in addition to the tighter $\chi^2$ constraint.

\begin{figure}
 \includegraphics[angle=-90,width=84mm]{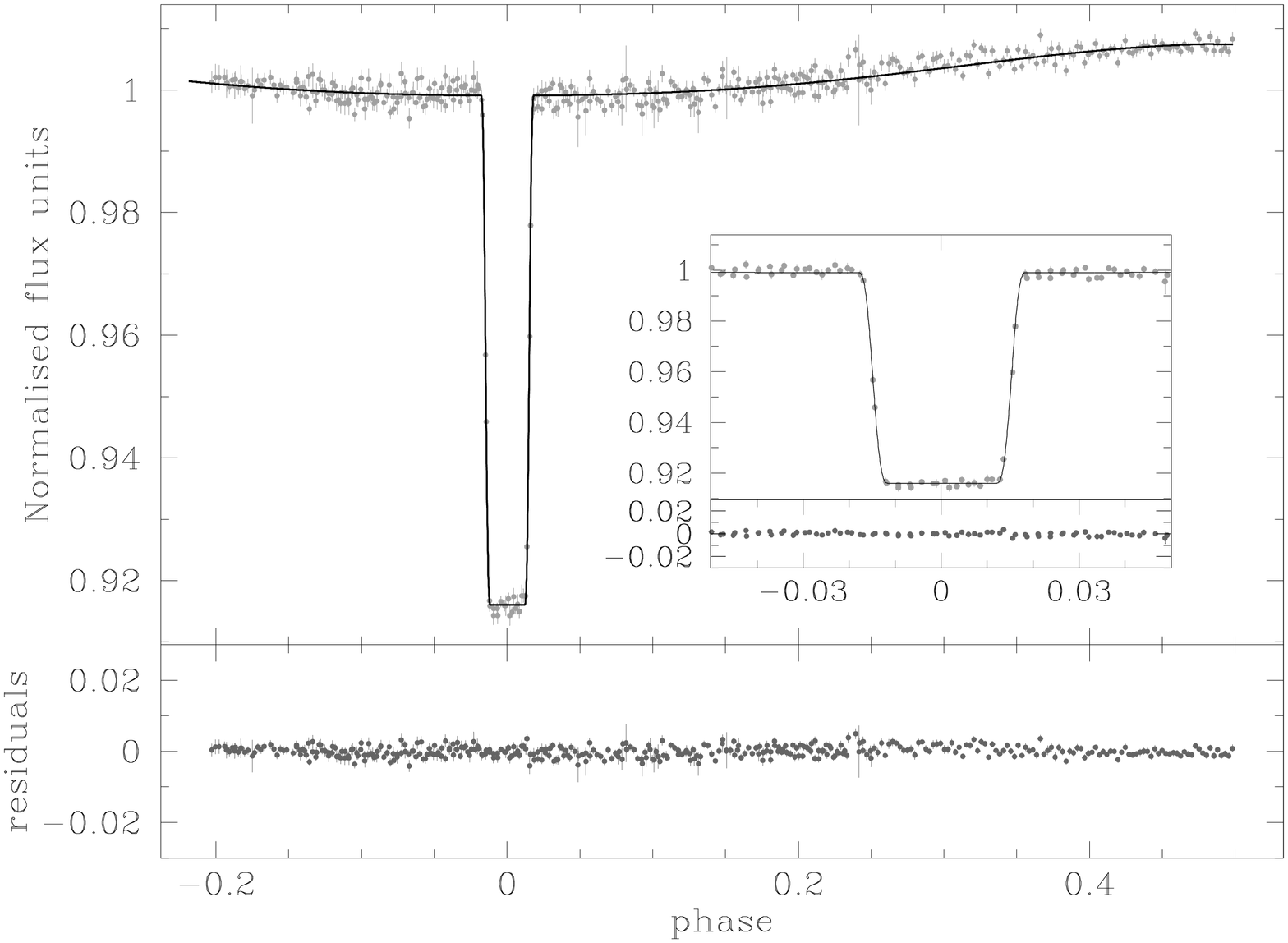}
  \caption{Model fit to the CA2.2 $I$-band light curve of SDSS\,0110+1326, for
    $\Mwd=0.59\,\Msun$, $\Rwd=0.015\,\Rsun$, $\Msec=0.28\,\Msun$,
    $\Rsec=0.32\,\Rsun$ $i=85^{\circ}$, and
    $K_\mathrm{sec}=197\,\kms$. The model radii agree with the
    theoretically predicted ones within $15\%$, for both the white
    dwarf and the secondary star. The residuals from the fit are shown
    at the bottom of the panel. Inset panel: data points and model fit
    focused around the eclipse.}
  \label{0110fit} 
\end{figure}

As an illustration of the quality of the light curve fits achieved,
Fig.\,\ref{0110fit} shows a fit with $\Mwd=0.59\,\Msun$,
$\Rwd=0.015\,\Rsun$, $\Msec=0.28\,\Msun$,
$\Rsec=0.32\,\Rsun$, $i=85^{\circ}$ and $K_\mathrm{sec}=197\,\kms$.
The mass of the secondary corresponds to a
spectral type of $\mathrm{Sp}(2)=\mathrm{M}4.5$. We stress, however,
that we do not adopt this model as a final solution for the parameters
of SDSS\,0110+1326, but rather consider it as a physically plausible
set of parameters.

Taking the $1\sigma$ confidence levels of the light curve fits at face
value, they imply a white dwarf mass of $\Mwdl=0.59-0.7\,\Msun$ and
a secondary mass of $\Msecl=0.28-0.38\,\Msun$. 

For the white dwarf, these numbers are in contrast with the results
from the spectral decomposition, $\Mwds=0.47\pm0.02\,\Msun$. The
errors in the spectroscopic white dwarf parameters are purely of
statistical nature. However, even assuming a reasonably large
systematic error of 0.2\,dex in $\log g$ (implying
$\Mwds=0.39-0.55\,\Msun$) the white dwarf masses implied by the light
curve fit and the spectroscopic fit appear to be inconsistent.

Regarding the secondary, adopting $\Msecl=0.28-0.38\,\Msun$ and
the $\mathrm{Sp}(2)-\Msec$ relation of
\citealt{rebassa-mansergasetal07-1} suggests a spectral type of
$\mathrm{Sp}(2)_\mathrm{lcfit}=\mathrm{M}3-\mathrm{M}4.5$, which is
consistent with the results of the spectral decomposition, 
$\mathrm{Sp}(2)_\mathrm{spfit}=\mathrm{M}4\pm1$.

While our analysis of the light curve fits demonstrates that the data
can be modelled well, it is clear that the lack of a reliable value
for $K_\mathrm{sec}$ prevents the choice of the physically most
meaningful solution.  Consequently, a definite determination of the
stellar parameters of SDSS\,0110+1326 is not possible on the basis of
the currently available data, and we adopt as a first cautious
estimate the values of our spectral decomposition fit. As mentioned
in Sect.\,\ref{s-lcmodels0110}, a physically motivated
$K$-correction can be modelled once a larger set of spectra covering
the entire binary orbit are available. 

%-------------------------------------------------------------------------------------------------------%
%-------------------------------------------------------------------------------------------------------%

\section{Post common envelope evolution}
\label{s-evolution}

Considering their short orbital periods, the four new eclipsing WDMS
binaries discussed in this paper must have formed through common
envelope evolution \citep{paczynski76-1, webbink07-1}.  Post common
envelope binaries (PCEBs) consisting of a white dwarf and a main
sequence star evolve towards shorter orbital periods by angular
momentum loss through gravitational radiation and magnetic braking
until they enter the semi-detached cataclysmic variable
configuration. As shown by \citet{schreiber+gaensicke03-1}, if the
binary and stellar parameters are known, it is  possible to
reconstruct the past and predict the future evolution of PCEBs for
a given angular momentum loss prescription. Here, we assume classical
disrupted magnetic braking \citep{rappaportetal83-1}, i.e. magnetic
braking is supposed to be much more efficient than gravitational
radiation but only present as long as the secondary star has a
radiative core.  For fully convective stars gravitational radiation is
assumed to be the only angular momentum loss mechanism, which applies
to all four systems.  The current ages of the PCEBs are determined by
interpolating the cooling tracks from
\citet{woodetal95-1}. Table\,\ref{t-evol} lists the evolutionary
parameters our four eclipsing WDMS binaries.

\begin{table}
\centering
\caption{Reconstructed and predicted orbital periods as well as expected 
evolutionary times for the four new eclipsing WDMS binaries. 
The cooling age of the white dwarfs is $t_{\mathrm{cool}}$ while 
$P_{\mathrm{CE}}$ and $P_{\mathrm{sd}}$ are the orbital periods at the end of
the common envelope phase and at the onset of mass transfer, respectively. The
time still needed to enter the semi-detached CV configuration is 
$t_{\mathrm{sd}}$.
}
\label{t-evol}
\begin{tabular}{lccccc}
\hline SDSS\,J & $t_{\mathrm{cool}}$ [yr]  & $P_{\mathrm{sd}}$ [d]&
$P_{\mathrm{CE}}$ [d] &
 $t_{\mathrm{sd}}$ [yr]\\ 
\hline 
0110+1326 & 1.41e+07 & 0.097& 0.332  & 1.78e+10 \\
1548+4057 & 3.91e+08 & 0.069 & 0.191 & 4.23e+09 \\
0303+0050 & 3.23e+09 & 0.097 & 0.228 & 5.95e+08 \\
1435+3733 & 2.53e+08 & 0.088 & 0.132 & 9.50e+08 \\
\hline
\end{tabular}
\end{table}
  
\begin{figure}
 \includegraphics[angle=270, width=84mm]{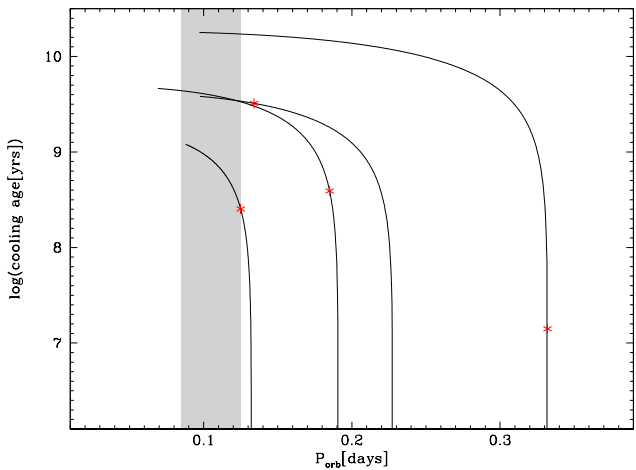}
  \caption{Evolutionary post common envelope tracks for our four
    eclipsing binaries. From left to right: SDSS\,1435+3733,
    SDSS\,1548+4057, SDSS\,0303+0054 and SDSS\,0110+1326. The current position 
    of the systems is indicated by an asterisk. We assumed only gravitational wave radiation
    as angular momentum loss agent as all four  systems are thought to
    contain fully convective secondary stars. The orbital period gap
    of CVs is highlighted in gray~--~all four systems will start mass
    transfer in, or shortly below the gap. 
  \label{f-evol}}
\end{figure}

The calculated evolutionary tracks of the four systems are shown in
Fig.\,\ref{f-evol}. The present position of each binary is indicated
by an asterisk. The grey-shaded region indicates the $2-3$\,h orbital
period gap, i.e. the orbital period range where only a small number of
cataclysmic variables (CVs) has been found.  Given that the four WDMS
binaries contain low-mass secondary stars, all of them are
expected to start mass transfer in or below the period gap. This
underlines the fact first noticed by \citet{schreiber+gaensicke03-1}
that only very few progenitors of long-period CVs are known.
SDSS\,0303+0050 has passed most of its PCEB lifetime and its current
orbital period differs significantly from the reconstructed value at
the end of the CE phase. The other three systems are rather young
PCEBs with a current orbital period close to $P_{\mathrm{CE}}$.

\section{Conclusions}
\label{s-conclusions}

\begin{table*}
\centering \setlength{\tabcolsep}{0.7ex}
\begin{minipage}{185mm}
\caption{Physical parameters of the known eclipsing white dwarf main
  sequence binaries.  A dash sign means that no corresponding value
  was given by the respective authors. Errors on the parameters are
  also shown, if quoted by the authors. } 
\label{allwdms}
\begin{tabular}{@{}lccccccccccl@{}}
\hline System & $P_{\mathrm{orb}}$ [d] & \Mwd [$\Msun$] & \Rwd [$\Rsun$] & \Twd [K] & log$g$ & \Msec [$\Msun$] & \Rsec [$\Rsun$] & $K_\mathrm{WD}$ [\kms] & $K_\mathrm{sec}$ [\kms] & Sp2 & Ref.\\ 
\hline 
V471\,Tau        & 0.521 & 0.79 & 0.01 & 32000 & - & 0.8 & 0.85 & - & 150.7$\pm1.2$ & K2 & 1a \\
                 & 0.521 & 0.84$\pm0.05$ & 0.0107$\pm0.0007$ & 34500$\pm1000$ & 8.31$\pm0.06$ & 0.93$\pm0.07$ & 0.96$\pm0.04$ & 163.6$\pm3.5$ & 148.46$\pm0.56$ & K2 & 1b \\
RXJ2130.6+4710   & 0.521 & 0.554$\pm0.017$ & 0.0137$\pm0.0014$ & 18000$\pm1000$ & 7.93\tiny{$^{+0.07}_{-0.09}$} & 0.555$\pm0.023$ & 0.534$\pm0.017$ & 136.5$\pm3.8$ & 136.4$\pm0.8$ & M3.5-M4 & 2\\
DE\,CVn          & 0.364 & 0.51\tiny{$^{+0.06}_{-0.02}$} & 0.0136\tiny{$^{+0.0008}_{-0.0002}$} & 8000$\pm1000$ & 7.5 & 0.41$\pm0.06$ & 0.37\tiny{$^{+0.06}_{-0.007}$} & - & 166$\pm4$ & M3V & 3 \\ 
GK\,Vir          & 0.344 & 0.7$\pm0.3$ & - & 50000 & - & - & - & - & - & M0-M6 & 4a \\
                 & 0.344 & 0.51$\pm0.04$ & 0.016 & 48800$\pm1200$ & 7.7$\pm0.11$ & 0.1 & 0.15 & - & - & M3-M5 & 4b\\
SDSS\,1212--0123 & 0.335 & 0.46-0.48 & 0.016-0.018 & $17700\pm300$ & 7.5-7.7 & 0.26-0.29 & 0.28-0.31 & - & $181\pm3$ & M4$\pm1$ & 5 \\
SDSS\,0110+1326  & 0.332 & 0.47$\pm0.2$ & 0.0163-0.0175 & 25900$\pm427$ & 7.65$\pm0.05$ & 0.255-0.38 & 0.262-0.36 & - & 180-200 & M3-M5 & 6 \\ 
RR\,Cae          & 0.303 & 0.365 & 0.0162 & 7000 & - & 0.089 & 0.134 & 47.9$^{*}$ & 195.8$^{*}$ & M6 & 7a \\
                 & 0.303 & 0.467 & 0.0152 & 7000 & - & 0.095 & 0.189 & - & 190$\pm9$ & M6 & 7b \\ 
                 & 0.303 & 0.44$\pm0.022$ & 0.015$\pm0.0004$ & 7540$\pm175$ & 7.61-7.78 & 0.183$\pm0.013$ & 0.188-0.23 & 79.3$\pm3$ & 190.2$\pm3.5$ & M4 & 7c\\ 
SDSS\,1548+4057  & 0.185 & 0.614-0.678 & 0.0107-0.0116 & 11700$\pm820$ & 8.02$\pm0.44$ & 0.146-0.201 & 0.166-0.196 & - & 274.7 & M5.5-M6.5 & 6 \\
EC13471-1258     & 0.150 & 0.77$\pm0.04$ & - & 14085$\pm100$ & 8.25$\pm0.05$ & 0.58$\pm0.05$ & 0.42$\pm0.08$ & - & 241$\pm8.1$ & M2-M4 & 8a \\
                 & 0.150 & 0.78$\pm0.04$ & 0.011$\pm0.01$ & 14220$\pm300$ & 8.34$\pm0.2$ & 0.43$\pm0.04$ & 0.42$\pm0.02$ & 138$\pm10$ & 266$\pm5$ & M3.5-M4 & 8b\\ 
CSS\,080502$^\dag$ & 0.149 & $0.35\pm0.04$ & $0.02\pm0.002$ & $17505\pm516$ & $7.38\pm0.12$ & 0.32 & 0.33 & - & - & M4 & 9,6\\
SDSS\,0303+0054  & 0.134 & 0.878-0.946 & 0.0085-0.0093 & $<\,$8000 & 8.4-8.6 & 0.224-0.282 & 0.246-0.27 & - & 339.7 & M4-M5 & 6 \\ 
NN\,Ser          & 0.130 & 0.57$\pm0.04$ & 0.017-0.021 & 55000$\pm8000$ & 7-8 & 0.10-0.14 & 0.15-0.18 & - & 310$\pm10$ & M4.7-M6.1 & 10a,b \\
                 & 0.130 & 0.54$\pm0.05$ & 0.0189$\pm0.001$ & 57000$\pm3000$ & 7.6$\pm0.1$ & 0.15$\pm0.008$ & 0.174$\pm0.009$ & 80.4$\pm4.1$ & 289.3$\pm12.9$ & M4.5-M5 & 10c \\ 
SDSS\,1435+3733  & 0.125 & - & - & - & - & 0.15-0.35 & 0.17-0.32 & - & - & M4-M6 & 11 \\ 
	 	 & 0.125 & 0.48-0.53 & 0.0144-0.0153 & 12500$\pm488$ & 7.62$\pm0.12$ & 0.19-0.246 & 0.218-0.244 & - & 260.9 & M4-M5 & 6 \\
CSS080408$^\ddag$ & - & $0.40\pm0.05$ & $0.022\pm0.003$ & $32595\pm675$ & $7.36\pm0.17$ & 0.26 & 0.27 & - & - & M5 & 9,6 \\
\hline
\multicolumn{12}{p{\textwidth}}{References:
           (1a)\,\citealt{young+nelson72-1};
           (1b)\,\citealt{obrienetal01-1};
           (2)\,\citealt{maxtedetal04-1};
           (3)\,\citealt{vandenbesselaaretal07-1};
           (4a)\,\citealt{greenetal78-1}; 
           (4b)\,\citealt{fulbrightetal93-1};
           (5)\,\citealt{nebotgomez-moranetal08-1}; 
           (6)\,this work; 
           (7a)\,\citealt{bruch+diaz98-1};
           (7b)\,\citealt{bruch99-1}; 
           (7c)\,\citealt{maxtedetal07-1};
           (8a)\,\citealt{kawkaetal02-1}; 
           (8b)\,\citealt{odonoghueetal03-1};
           (9)\,\citealt{drakeetal08-1};
	   (10a)\,\citealt{wood+marsh91-1};	
           (10b)\,\citealt{catalanetal94-1};
           (10c)\,\citealt{haefneretal04-1};
           (11)\,\citealt{steinfadtetal08-1}} \\ 
\hline
\multicolumn{12}{p{\textwidth}}{Notes: $^*$ Calculated from $K\sin i$
  provided by the authors; $^\dag$
  CSS080502:090812+060421\,=\,SDSS\,J090812.04+060421.2, we determined
  the white dwarf 
  and secondary star paremeters from  decomposing/fitting the SDSS
  spectrum (see Sect.\,\ref{s-spectralfit} and
  \citealt{rebassa-mansergasetal07-1}), these parameters should be taken
  as guidance as no light curve modelling has been done so far; $^\ddag$
  CSS080408:142355+240925\,=\,SDSSJ142355.06+240924.3, also found by
  \citet{rebassa-mansergasetal08-2}, white dwarf and secondary star
  parameters determined as for CSS080502.}\\ 
\hline
\end{tabular}
\end{minipage}
\end{table*}

Since \citet{nelson+young70-1} reported the discovery of V471\,Tau, as
the first eclipsing white dwarf main sequence binary, only 7
additional similar systems have been discovered, the latest being
SDSS\,1435+3733 found by \citet{steinfadtetal08-1}. Here, we presented
follow-up observations of SDSS\,1435+3733, which we independently
identified as an eclipsing WDMS binary, and of three new discoveries:
SDSS\,0110+1326, SDSS\,0303+0050, and SDSS\,1548+4057. A fifth system
has just been announced by our team \citep{nebotgomez-moranetal08-1},
and two eclipsing WDMS binaries were identified by
\citet{drakeetal08-1} in the Catalina Real-Time Transient Survey.
Table\,\ref{allwdms} lists the basic physical parameters of the
fourteen eclipsing white dwarf main sequence binaries currently
known. 

Inspection of Table\,\ref{allwdms} shows that the orbital periods of
the known eclipsing binaries range from three to 12 hours, with three
of the systems presented here settling in at the short-period end.

The white dwarf masses cover a range $\sim0.44-0.9\,\Msun$, with an
average of $<\Mwd>=0.57\pm0.16\,\Msun$, which is only slightly lower
than the average mass of single white dwarfs \citep{liebertetal05-1,
  finleyetal97-1, koesteretal79-1}. This finding is somewhat
surprising, as all these eclipsing WDMS binaries have undergone a common
envelope evolution, which is expected to cut short the evolution of
the primary star and thereby to produce a significant number of
low-mass He-core white dwarfs.  Our estimates for the white dwarf
masses in SDSS\,0110+1326 and SDSS\,1435+3733 suggest that they
contain He-core white dwarfs. The white dwarf in SDSS\,1548+4057 is
consistent with the average mass of single white
dwarfs. SDSS\,0303+0054 contains a fairly massive white dwarf,
potentially the most massive one in an eclipsing WDMS binary.

The secondary star masses are concentrated at very low masses, the
majority of the fourteen systems in Table\,\ref{allwdms} have
$\Msec<0.6$\,\Msun, and $\sim9$ systems (including the four stars
discussed here) have $\Msec\la0.3\,\Msun$. This is a range where very
few low-mass stars in eclipsing binaries are known \citep{ribas06-1},
making the secondary stars in eclipsing WDMS binaries very valuable
candidates for filling in the empirical mass-radius relation of low
mass stars. This aspect will be discussed in a forthcoming paper.

The distribution of secondary star masses among the eclipsing PCEBs in Table~\ref{allwdms} is subject to similar selection effects as the whole sample of known PCEBs. \citet{schreiber+gaensicke03-1} found a strong bias towards late-type secondary stars in the pre-SDSS PCEB population, which arose as most of those PCEBs were identified in blue-colour surveys. The currently emerging population of SDSS PCEBs \citep{silvestrietal06-1, rebassa-mansergasetal08-2} should be less biased as the $ugriz$\,--\,colour-space allows to identify WDMS binaries with a wide range of secondary spectral types and white dwarf temperatures.

Future work will need to be carried out on two fronts. On one hand,
more accurate determination of the masses and radii will have to be
made to unlock the potential that WDMS binaries hold for
testing/constraining the mass-radius relation of both white dwarfs and low
mass stars. The key to this improvement are high-quality light curves
that fully resolve the white dwarf ingress/egress, and, if possible,
the secondary eclipse of the M-dwarf by the white dwarf. In addition,
accurate $K_\mathrm{WD}$ velocities are needed, which will be most reliably
obtained from ultraviolet intermediate resolution spectroscopy. On the
other hand, it is clear that the potential of SDSS in leading to the
identification of additional eclipsing WDMS binaries has not been
exhausted. Additional systems will be identified by the forthcoming
time-domain surveys, and hence growing the class of eclipsing WDMS
binaries to several dozen seems entirely feasible. 

\section*{Acknowledgements}
We thank Justin Steinfadt for
providing his light curve data on SDSS\,1435+3733 and the referee, Pierre Maxted, for his swift
report and constructive suggestions.
Based in part on observations made with the William Herschel Telescope
operated on the island of La Palma by the Isaac Newton Group in the
Spanish Observatorio del Roque de los Muchachos of the Instituto de
Astrof\'isica de Canarias; on observations made with the Nordic
Optical Telescope, operated on the island of La Palma jointly by
Denmark, Finland, Iceland, Norway, and Sweden, in the Spanish
Observatorio del Roque de los Muchachos of the Instituto de
Astrof\'isica de Canarias; on observations collected at the Centro
Astron\'omico Hispano Alem\'an (CAHA) at Calar Alto, operated jointly
by the Max-Planck Institut f\"ur Astronomie and the Instituto de
Astrof\'isica de Andaluc\'ia (CSIC); on observations made with the
Mercator Telescope, operated on the island of La Palma by the Flemish
Community, at the Spanish Observatorio del Roque de los Muchachos of
the Instituto de Astrof\'isica de Canarias; and on observations made
with the IAC80 telescope operated by the Instituto de Astrof\'isica de
Canarias in the Observatorio del Teide.

\bibliographystyle{mn_new} 
\bibliography{aamnem99,aabib,proceedings,submitted}

\bsp

\label{lastpage}

\end{document}